\documentclass[aip,
amsmath,amssymb,
reprint,%
]{revtex4-1}

\usepackage[colorlinks=true,citecolor=blue,urlcolor=blue,linkcolor=blue,bookmarksopen]{hyperref}
\usepackage{graphicx}
\usepackage{dcolumn}
\usepackage{bm}

\usepackage[utf8]{inputenc}
\usepackage[T1]{fontenc}
\usepackage{mathptmx}
\usepackage{MnSymbol}
\newcommand{\Vpol}{\mathrel{\updownarrow}}
\newcommand{\Hpol}{\mathrel{\leftrightarrow}}
\newcommand{\Dpol}{\mathrel{\text{$\nearrow$\llap{$\swarrow$}}}}
\newcommand{\Apol}{\mathrel{\text{$\nwarrow$\llap{$\searrow$}}}}
\renewcommand{\vec}[1]{\boldsymbol{#1}}

\begin{document}
	
	\preprint{AIP/123-QED}
	
	\title{Realization of room temperature polaritonic vortex in momentum space with hybrid Perovskite metasurface}
	
	
	\author{Nguyen Ha My Dang}
	\affiliation{Univ Lyon, Ecole Centrale de Lyon, CNRS, INSA Lyon, Université Claude Bernard Lyon 1, CPE Lyon, CNRS, INL, UMR5270, 69130 Ecully, France} 
	\author{Simone Zanotti}
	\affiliation{Dipartimento di Fisica, Universit\`{a} di Pavia, via Bassi 6, I-27100 Pavia, Italy}
	\author{Emmanuel Drouard}
	\affiliation{Univ Lyon, Ecole Centrale de Lyon, CNRS, INSA Lyon, Université Claude Bernard Lyon 1, CPE Lyon, CNRS, INL, UMR5270, 69130 Ecully, France} 
	\author{C\'{e}line Chevalier}
	\affiliation{Univ Lyon, Ecole Centrale de Lyon, CNRS, INSA Lyon, Université Claude Bernard Lyon 1, CPE Lyon, CNRS, INL, UMR5270, 69130 Ecully, France} 
	\author{Ga\"{e}lle Tripp\'e-Allard}
	\affiliation{Universit\'{e} Paris-Saclay, ENS Paris-Saclay, CNRS, CentraleSup\'{e}lec, LuMIn, Gif-sur-Yvette, France}
	\author{Mohamed Amara}
	\affiliation{Univ Lyon, Ecole Centrale de Lyon, CNRS, INSA Lyon, Université Claude Bernard Lyon 1, CPE Lyon, CNRS, INL, UMR5270, 69130 Ecully, France} 
	\author{Emmanuelle Deleporte}
	\affiliation{Universit\'{e} Paris-Saclay, ENS Paris-Saclay, CNRS, CentraleSup\'{e}lec, LuMIn, Gif-sur-Yvette, France}
	\author{Vincenzo Ardizzone}
	\affiliation{CNR Nanotec, Institute of Nanotechnology, via Monteroni, 73100, Lecce}
	\author{Daniele Sanvitto}
	\affiliation{CNR Nanotec, Institute of Nanotechnology, via Monteroni, 73100, Lecce}
	\author{Lucio Claudio Andreani}
	\affiliation{Dipartimento di Fisica, Universit\`{a} di Pavia, via Bassi 6, I-27100 Pavia, Italy}
	\author{Christian Seassal}
	\affiliation{Univ Lyon, Ecole Centrale de Lyon, CNRS, INSA Lyon, Université Claude Bernard Lyon 1, CPE Lyon, CNRS, INL, UMR5270, 69130 Ecully, France} 
	\author{Dario Gerace}
	\affiliation{Dipartimento di Fisica, Universit\`{a} di Pavia, via Bassi 6, I-27100 Pavia, Italy}
	\author{Hai Son Nguyen}
	\email{hai-son.nguyen@ec-lyon.fr}
	\affiliation{Univ Lyon, Ecole Centrale de Lyon, CNRS, INSA Lyon, Université Claude Bernard Lyon 1, CPE Lyon, CNRS, INL, UMR5270, 69130 Ecully, France} 
	\affiliation{IUF, Universit\'e de France}
	
	\date{\today}
	
	\begin{abstract}
		Exciton-polaritons are  mixed light-matter excitations that result from the strong coupling regime between an active excitonic material and photonic resonances. Harnessing these hybrid excitations provides a rich playground to explore fascinating fundamental features, such as out-of-equilibrium Bose-Einstein condensation and quantum fluids of light, as well as novel mechanisms to be exploited in optoelectronic devices. Here, we investigate experimentally the formation of exciton-polaritons arising from the mixing between hybrid inorganic-organic perovskite excitons and an optical Bound state In a Continuum (BIC) of a subwavelength-scale metasurface, at room temperature. These polaritonic eigenmodes, hereby called polariton BICs (pol-BICs) are revealed in both reflectivity, resonant scattering, and photoluminescence measurements. Although pol-BICs only exhibit a finite quality factor that is bounded by the non-radiative losses of the excitonic component, they fully inherit BIC peculiar features: a full uncoupling from the radiative continuum in the vertical direction, which is associated to a locally vanishing farfield radiation in  momentum space. Most importantly, our experimental results confirm that the topological nature of the photonic BIC is perfectly transferred to the pol-BIC. This is evidenced with the observation of a polarization vortex in the farfield of polaritonic emission.  Our results pave the way to engineer BIC physics of interacting bosons, as well as novel room temperature polaritonic devices.  
	\end{abstract}
	
	\maketitle

	\section{\label{sec:level1}Introduction}
	Bound state in the Continuum (BICs) are peculiar localized states that are forbidden to radiate despite lying in a continuum of propagating waves.  Once regarded as an `exotic' quantum mechanical effect\,\cite{VonNeumann1929}, the origin of BICs is nowadays fully unravelled as a particular solution of wave equations, which has led to their exploitation in other fields where it is straightforwardly attributed to destructive interference mechanisms or symmetry mismatches\,\cite{Hsu2016}. The most active playground of BIC physics is in contemporary Photonics\,\cite{Azzam2021} since most of optoelectronic devices rely on resonances and their coupling mechanisms with environment. Indeed, the trapping of light through photonic BICs is a salient feature to enhance different light-matter interaction mechanisms, leading to various applications in microlasers\cite{Kodigala2017,Jin2019,Wu2020,BICSwitch,Hwang2021}, sensoring\cite{Romano:18}, optical switch\cite{BICSwitch} and nonlinear optics\cite{BIC_nonlinear,Minkov:19,BICnonlinear2,Houdre2020}. Moreover, on the fundamental side, photonic BICs in periodic lattices are pinned to singularities of farfield polarization vortex and can be considered as topological charges of non-Hermitian systems\cite{Zhen2014,Doeleman2018,Zhang2018,Wang2020}. This topological nature, together with modern technological feasibility to tailor photonic materials, makes photonic BICs a fruitful platform to engineer polarization singularities of open photonic systems\cite{yin2019,Jin2019,Ye2020,Yoda2020}. 
	
	Another prominent area of investigation in modern Optics is represented by exciton-polaritons, elementary excitations arising from the strong coupling regime between confined light and semiconductor excitons\cite{Claude1994}. As hybridized eigenmodes, these bosonic quasiparticles inherit original features of both photonic resonances and  excitonic ones. During the last decades, exciton-polaritons have become a standard platform to explore exotic phenomena of out-of-equilibrium Bose Einstein condensations\cite{Kasprzak2006} and quantum fluid properties of electromagnetic fields\cite{QuantumFLuidsOfLight}, as well as novel concepts for all-optical devices. While most of the milestones of polaritonic physics have been achieved with excitonic material operating in cryogenic temperatures ($\sim$ 10\,K), recent progress in material science has promoted many polaritonic demonstrations at room temperature with a wide variety of high bandgap semiconductors, ranging from GaN\cite{Christopoulos2007,Daskalakis2013}, ZnO\cite{Franke_2012}, to organic semiconductors\cite{Lidzey1998,Plumhof2014} and transition metal dichalcogenide monolayers\cite{Liu2014,Grosso2017}, as well as perovskite materials\cite{Fujita1998,Takada2003,lanty2008strong,Nguyen2014,Su2017,Fieramosca2018,Su2020,Dang2020}.
	
	Very recently, the strong coupling regime between photonic BICs and excitonic resonances has been theoretically suggested,\cite{Koshelev2018,Lu:20,Xie2021,Ani2021} with two experimental demonstrations\cite{Kravtsov2020,ardizzone2021polariton}. The result of such a coupling is the formation of polariton-BICs (pol-BICs):  hybrid excitations that are completely decoupled from the radiative continuum. This scenario is quite different from the text-book architecture in which the strong coupling regime is engineered by embedding quantum wells at the antinode positions of planar microcavities, which always exhibit substantial nonradiative losses. As state-of-the-art, non-linear behaviors of pol-BICs\cite{Kravtsov2020} and the use of pol-BICs to trigger Bose-Einstein Condensation  have been recently demonstrated\cite{ardizzone2021polariton}. However, most of these demonstrations are conducted at cryogenic temperatures. Therefore, making pol-BICs with room temperature operation would be the next important step for the development of polaritonic devices based on BIC concepts. 
	
	Perovskite material has appeared in the last few years as the ideal alternative candidate to study exciton-polaritons at room temperature.\cite{Su2021} Nowadays, polaritonic Bose-Einstein condensation in the perovskite platform can be engineered to reproduce many features that have been up to now only evidenced in cryogenic GaAs-based polaritons.\cite{Nguyen2014,Su2017,Su2020,Dang2020} Moreover, the possibility to obtain high quality perovskite layers via solution-based methods opens feasible and low-cost approaches to fabricate perovskite metasurfaces. In this approach, the tailoring of the real part (i.e. frequency) of polaritonic energy-momentum dispersion has been experimentally demonstrated with sub-wavelength perovskite metasurface, in which linear, parabolic and multi-valley dispersion characteristics have been reported.\cite{Dang2020} Most recently, it has been theoretically suggested that perovskite metasurfaces can be harnessed to engineer the imaginary part (i.e., losses) of the polaritonic energy-momentum dispersion at room temperature via BIC and Exceptional Point concepts.\cite{Lu:20,Ani2021} 
	
	In this work, we experimentally demonstrate the formation of pol-BICs in an excitonic metasurface made from hybrid 2D perovskites. The pol-BICs, revealed in both reflectivity, resonant scattering, and photoluminescence measurements, are robust at room temperature. They are originated from the strong coupling regime between a symmetry-protected BIC and the excitonic resonance of the perovskite material. Although the pol-BICs only exhibit finite quality factor, which is actually limited by the non-radiative losses from the excitonic component, they fully inherit the purely photonic BIC peculiar features: the decoupling from the radiative continuum, and the farfield emission vanishing at $\Gamma$ point. Remarkably, the pol-BICs also inherit the topological nature of the purely photonic BICs. This is demonstrated through the observation of a  polarization vortex in the farfield emission of pol-BIC modes. 
	
	\section{\label{sec:level2}Results and Discussion}
\begin{figure}[htb]
		\begin{center}
			\includegraphics[width=0.5\textwidth]{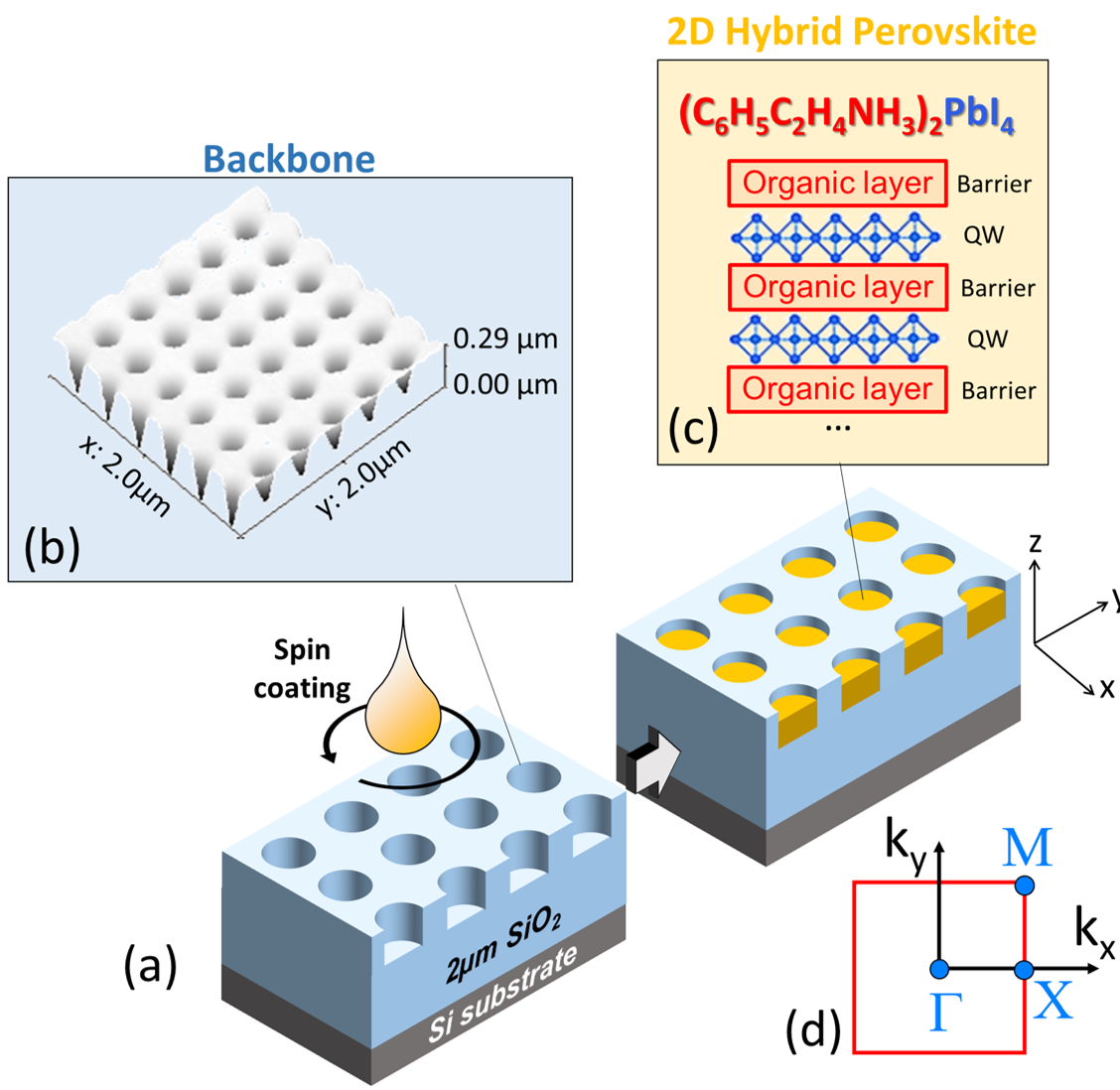}
		\end{center}
		\caption{\label{fig1}{a). Sketch of the fabrication method of the perovskite metasurface. The backbone consists of a 2\,$\mu$m-thick thermal SiO$_2$ on silicon substrate. It is pre-patterned into a square lattice of holes, with period $a=330$\, nm, diameter $d=132$\, nm and depth $t_{SiO2}=150$\, nm. The crystallized perovskite layer has a total thickness $t_{PEPI}=180$\, nm. Thus there is a residual 30\,nm-PEPI slab (not shown in the sketch). The final structure is encapsulated by 200\,nm-thick PMMA (not shown in the sketch).  b) Atomic Force Microscopy (AFM) image of the photonic backbone. c) The layered structure of PEPI.(d) The reciprocal space of the our metasurface with square Brillouin zone of size $2\pi/a$.}}
	\end{figure}

	Our excitonic material is a 2D layered hybrid  organic-inorganic perovskite, namely bi-(phenethyl\-ammonium) tetraiodoplumbate, known as PEPI, with chemical formula (C$_6$H$_5$C$_2$H$_4$NH$_3$)$_2$PbI$_4$. Thanks to the alternation between organic and inorganic monolayers (see Fig.~\ref{fig1}b), PEPI thin films behave like a multi-quantum well structure where excitons are confined within inorganic monolayers. PEPI is one of the most widespread perovskite material for polaritonic applications, thanks to its giant exciton binding energy ($\sim$ hundred meV at room temperature)\cite{Gauthron2010}. Here the fabrication of PEPI metasurface follows the same method previously reported (see Fig.~\ref{fig1}c for a simplified sketch).\cite{Dang2020} The PEPI solution is infiltrated inside air holes of pre-patterned silica backbone via spincoating, then followed by a thermal annealing to form a square lattice of crystallized PEPI nano-pillars. The final sample is encapsulated  with Poly-methyl methacrylate (PMMA) to protect PEPI against humidity.  
	\begin{figure*}[htb]
		\begin{center}
			\includegraphics[width=0.9\textwidth]{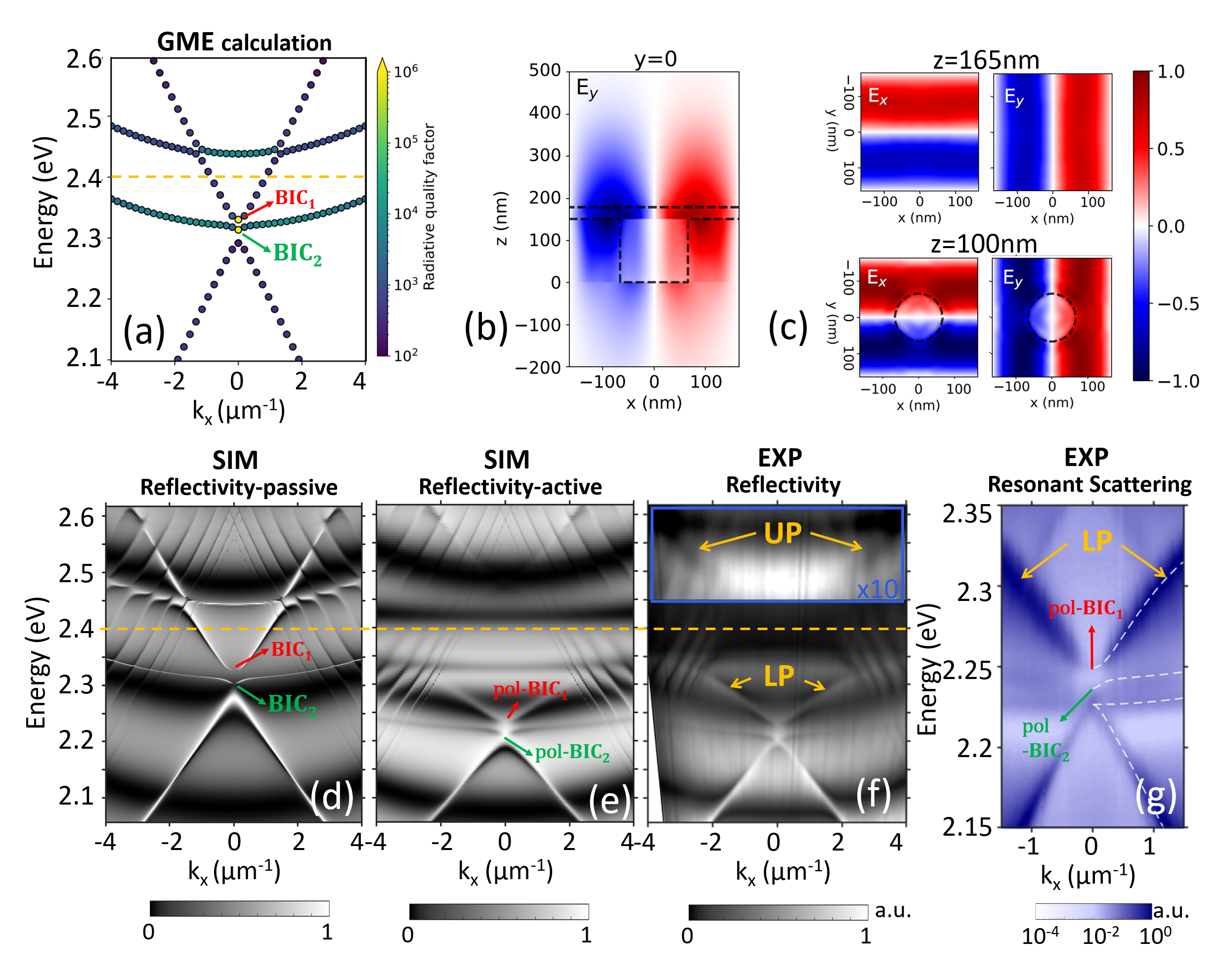}
		\end{center}
		\caption{\label{fig2}{(a) Energy-momentum dispersion calculated by GME for the passive metasurface by GME method for odd modes with respect to mirror symmetry for reflection against the $xz$ plane (at $y=0$); the Q-factor of each $k_x-$dependent mode is encoded in the color scale of each point (yellow corresponds to virtually infinite Q-factor, as determined from GME simulations). The dotted orange line indicates the exciton energy: $E_X=2.394$\,eV, and it is shown here for illustration purposes. (b) Spatial distribution of the electric field in the $xz$ plane within a single unit cell of the square lattice, as obtained for the BIC$_1$ mode, and (b) corresponding cross sections in the $xy$ plane for two different vertical coordinates, corresponding either to the patterned region (PEPI infiltrated hole) or the uniform PEPI region; we notice that the calculated field distribution for the BIC$_2$ is very similar. (d) Numerical simulations of the angle-resolved reflectivity spectra calculated for the passive structure (assuming the same parameters as the GME calculation), and (e) same simulation for the active structure (i.e., assuming a Lorentz oscillator model for the PEPI material, see methods), when excited by $\Vpol$ (i.e. $y$ ) polarized light; also here the dashed line indicates the bare excitonic resonance of PEPI. (f) Experimental result of the angle-resolved reflectivity of the PEPI metasurface when excited by $\Vpol$ polarized light. (g) Experimental result of the angle-resolved resonant scattering of the PEPI metasurface, zoomed in the vicinity of the two pol-BICs. The excitation is polarized along $\Dpol$ (i.e. 45 degrees) and the detection is analysed along $\Apol$ (i.e. -45 degrees). The dotted lines are guides to the eye to distinguish the 4 polaritonic modes. }}
	\end{figure*}
	
	Our excitonic metasurface is designed so that the exciton energy ($E_X=2.394$\,eV) is in the vicinity of the lowest Bloch resonances of the passive structure (PEPI is replaced by a dielectric material of constant refractive index $n=2.4$). In Fig.~\ref{fig2}a we report the energy-momentum dispersion and the corresponding quality factor (color-coded into the plot by points) of such Bloch resonances as calculated by Guided-Mode Expansion (GME) method,\cite{Andreani-Gerace2006} generalized to deal with multilayered and partially etched dielectric materials.\cite{Minkov2020} The dispersion is calculated along $\Gamma X$ direction (i.e. $k_x$, see Fig.~\ref{fig1}d). Only modes with an odd parity with respect to mirror symmetry for reflection through the $xz$ ($y=0$) plane are shown in Fig.~\ref{fig2}a (see supplementary material for the whole spectrum including even modes), which can be excited by transverse-electric polarized incident radiation. These results show that the lowest mode is a bright (low Q-factor) one at $\Gamma$ point, while two singly-degenerate BICs are present at slightly higher energy, here denoted BIC$_1$ and BIC$_2$. The theoretical electric field patterns of these two photonic BICs are reported in Fig.~\ref{fig2}b,c, showing both $E_x$ and $E_y$ components, both in vertical and planar cross sections. The odd parity of these spatial profiles for mirror symmetry through either $xz$ or $yz$ planes confirms that they are both symmetry-protected BICs, i.e., they are forbidden to couple to the radiative continuum due to their symmetry mismatch with radiative plane waves.  As a result of numerical simulations using the Rigorous Coupled Wave Analysis (RCWA) method, Fig.~\ref{fig2}d presents the calculated angle-resolved reflectivity of the passive structure when excited by $\Vpol$ (i.e. $y$) polarized light, that is s-polarized (transverse electric, TE) with respect to the $xz$ plane of incidence. These results show very good agreement with the GME calculations for odd modes, we clearly distinguish the two BICs corresponding to the local vanishings of reflectivity resonances at $k_x=0$. We also notice that only a singly degenerate leaky mode is shown in Fig.~\ref{fig2}d with a maximum slightly below 2.3 eV, which matches the result in Fig.~\ref{fig2}a. The other mode, degenerate in $\Gamma$, is of $\Hpol$ (i.e., $x$) polarization (clearly evidenced in the corresponding GME results for even modes and the corresponding RCWA simulations for p-polarized (transverse magnetic, TM) excitation with respect to the $xz$ incidence plane, shown in supplementary material). 
	
	We now investigate the coupling  between PEPI excitons and the previous Bloch resonances, giving rise to polaritonic modes.\cite{Gerace-Andreani2007}. A home-built setup for Fourier spectroscopy is employed to study the excitonic metasurface via angle-resolved and spectral-resolved experiments. The experimental results of angle-resolved reflectivity measurements (Fig.~\ref{fig2}f) and numerical ones from RCWA simulation (Fig.~\ref{fig2}e) are in perfect agreement. The strong coupling regime is evidently demonstrated in both experimental and numerical results, by comparison with the simulations of the passive structures (i.e., without considering the excitonic response) in Fig.~\ref{fig2}d. Indeed, all dispersion curves are strongly red-shifted from the bare exciton energy to form polaritonic branches. In particular, the anticrossings between the first lower polariton branch, here denoted LP, and the first upper polariton branch, here denoted UP, are clearly evidenced. The measured Rabi splitting is $\hbar\Omega\approx 200$\,meV, which is in good agreement with previous reports on polaritons in PEPI metasurface,\cite{Dang2020} as well as from the known expression for the radiation-matter coupling in confined systems in terms of the bulk material one\cite{StrongCoupling_book}, see also details in Methods.
	
	Since the polaritonic modes inherit the symmetry properties of their photonic component, the first two LP branches possess the same odd parity at $\Gamma$ point as the one of BIC$_1$ and BIC$_2$. Therefore, they must be symmetry-protected pol-BICs, denoted as pol-BIC$_1$ and pol-BIC$_2$, according to their photonic counterparts. To evidence their decoupling from the radiative electromagnetic continuum, a straightforward criterion is the vanishing of these resonances at $\Gamma$ point, i.e. at normal incidence with respect to the metasurface. Although these vanishing resonances are hinted in Figs.~\ref{fig2}e,f, it is a tricky observation because the reflectivity resonances corresponding to these guided modes have Fano-like  profiles superimposed on a shallow Fabry-Perot modulation deriving from the 2\,$\mu$m-thick SiO$_2$.\cite{Galli2005} We get rid of such a background by employing resonant scattering measurements,\cite{McCutcheon2005} in which the excitation and detection are in cross-polarization configuration (excitation in 45-degrees polarization direction, and analyzed in -45-degrees one). As shown in Fig.~\ref{fig2}g, we observe all four polariton lower branches, corresponding to the four Bloch resonances previously discussed. Notably, the resonant scattering results feature both $E_x$ and $E_y$ polarized modes. Indeed, the leaky branch of high curvature of $E_x$ polarization is observed in resonant scattering results, and not in reflectivity measurements of $E_y$ polarization. Most importantly, as expected, the resonant scattering vanishes locally at the two pol-BICs.  
	
	\begin{figure}[htb]
		\begin{center}
			\includegraphics[width=0.5\textwidth]{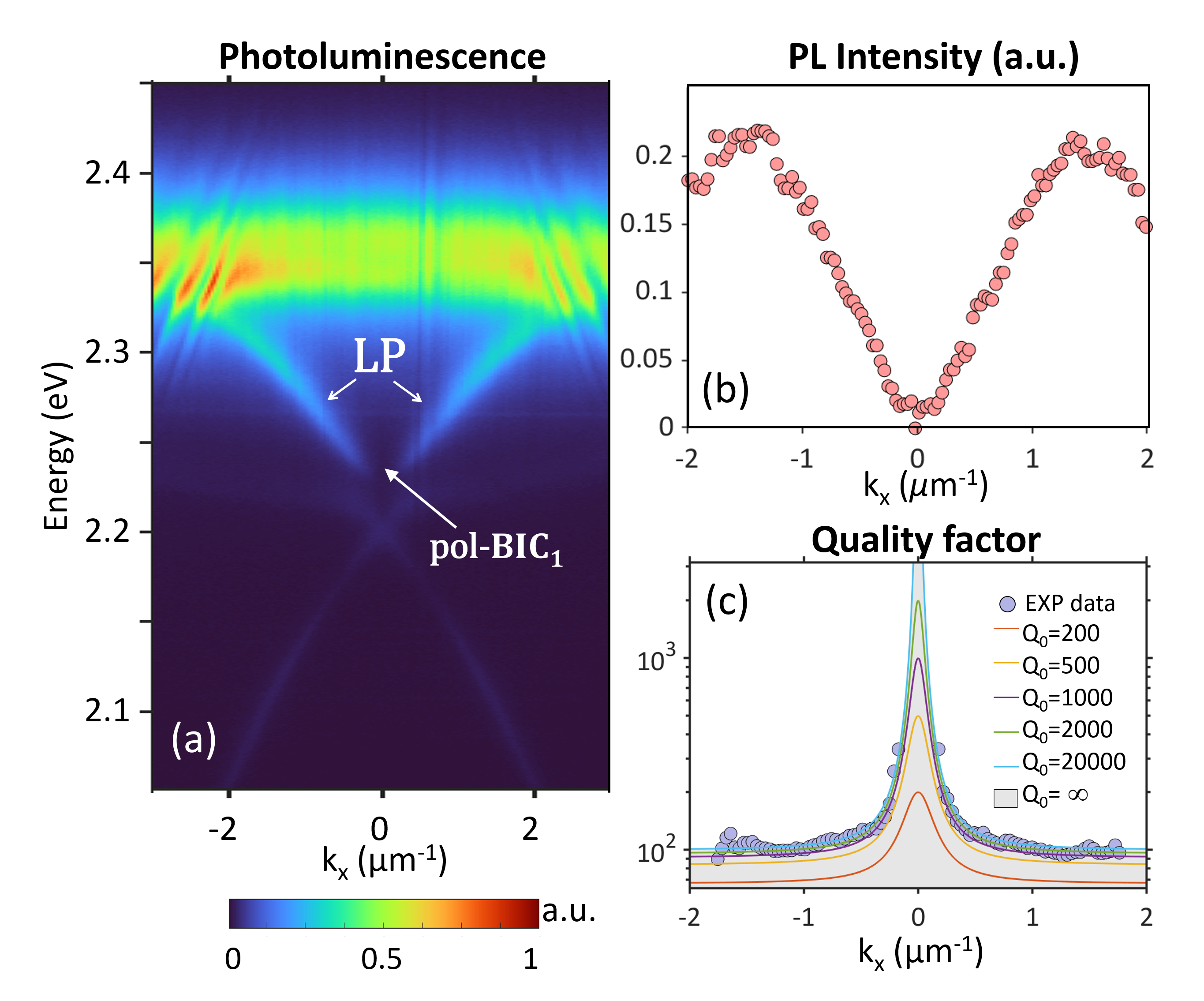}
		\end{center}
		\caption{\label{fig3}{(a) Angle-resolved photoluminescence results, analysed along $\Vpol$(i.e. $y$). (b) Angle-resolved emission intensity of the LP. (c) The quality factor of the LP extracted from the photoluminescence measurements. The blue circles are experimental data, the solid lines correspond to the model \eqref{eq:Q} with $Q_\infty=95$, $\alpha=5.9$\,$\mu$m$^2$ but using different values of $Q_0$. The gray shading area corresponds to $Q_0=\infty$.}}
	\end{figure}
	\begin{figure}[htb]
		\begin{center}
			\includegraphics[width=0.5\textwidth]{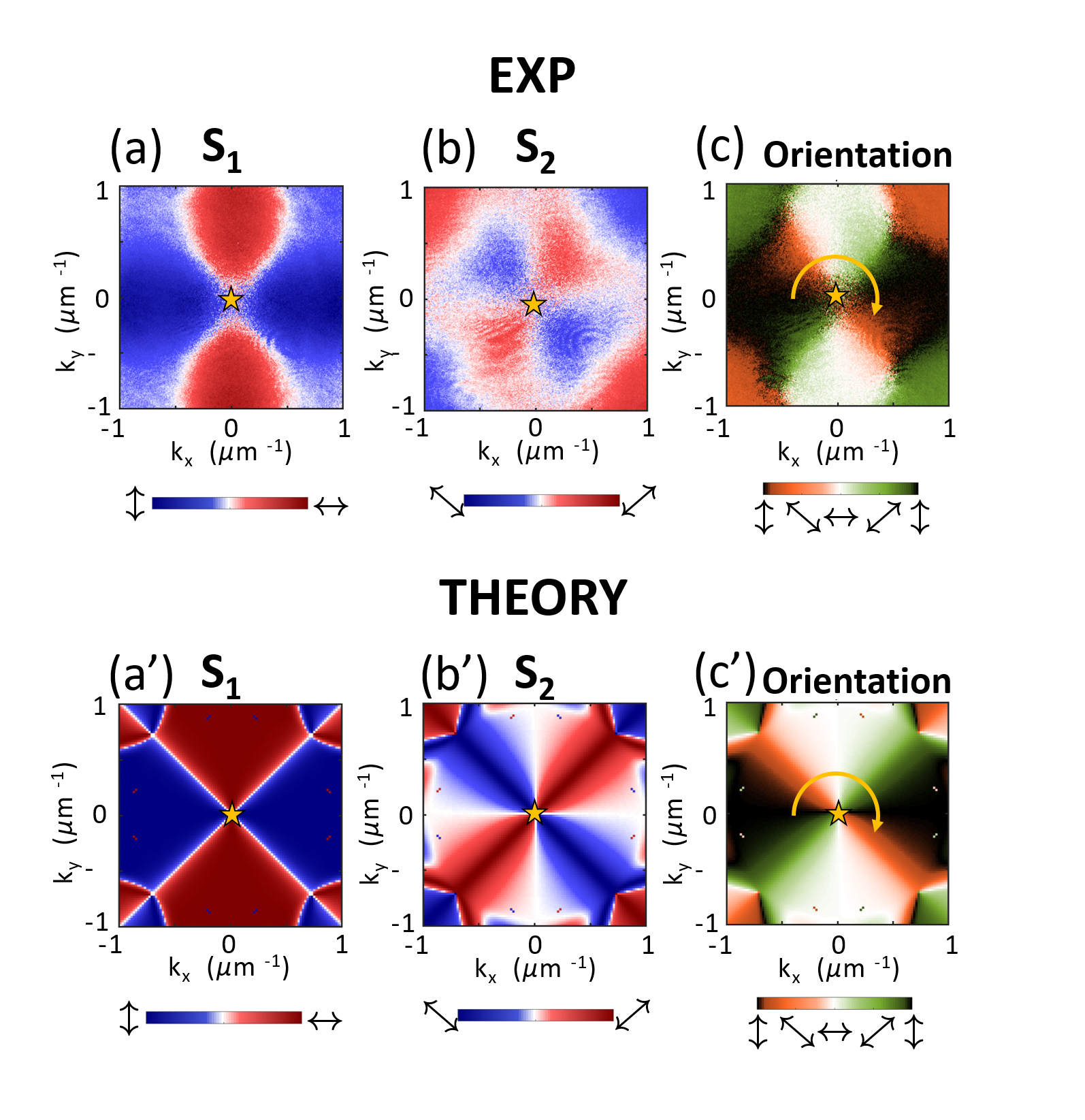}
		\end{center}
		\caption{\label{fig4} Experimental results for the mapping of the Stokes parameters $S_1$ (a), $S_2$ (b), and the polarization orientation corresponding to pol-BIC$_1$ (c). Theoretically calculated $S_1$ (a'), $S_2$ (b'), and polarization orientation (c'), as obtained from GME simulations of the passive structure and corresponding to the BIC$_1$ Bloch mode of Fig.~\ref{fig2}a. }
	\end{figure}
	
	The reflectivity and resonant scattering measurements show that the PEPI metasurface can host pol-BICs modes. To populate these polaritonic modes and study their farfield emission, we perform non-resonant pumping ($\lambda_{laser}=400$\,nm, 80\,MHz, 50\,ps) with the same Fourier spectroscopy setup. Angle-resolved photoluminescence results, shown in Fig.~\ref{fig3}a, indicate that only the LP branch is efficiently populated while other modes (the UP and photonic-like bands at low energy) are either not observed or weakly populated. This observation is in good agreement with polaritonic behaviours in high bandgap materials for which the photoluminescence emission is mainly visible for the first lower polariton branch. The farfield emission of the LP vanishes locally at $k_x=0$ where the pol-BIC$_{1}$ resides. Such effect, already shown in Fig.~\ref{fig3}a, is clearly evidenced in Fig.~\ref{fig3}b where the photoluminescence intensity of the LP is tracked and reported. We note that the excitonic fraction of the pol-BIC$_1$ amounts to 35$\%$ as estimated from a simple model of two coupled oscillators, in which we take half of the Rabi splitting as the radiation-matter coupling energy (off-diagonal element in the $2\time 2$ matrix, much larger than the oscillators linewidths) and the exciton-photon detuning estimated from the passive photonic BIC$_1$ mode (Fig.~\ref{fig2}a) indicating that it is indeed an hybrid exciton-photon entity.
	
	We now discuss on the use of a photonic BIC to enhance the quality factor of polaritonic modes. Since photonic BICs possess an ideally infinite quality factor, one may expect the same for pol-BICs. However, due to their hybrid exciton-photon nature, the quality factor of pol-BICs is balanced between the non-radiative losses of the excitonic component and the one from the photonic BIC\cite{Lu:20}. They are thus quasi-BICs, whose quality factor is limited by non-radiative losses. Moreover, the $C_4$ symmetry of the square-lattice imposes that the quality factor of photonic BICs decreases as $k_x^2$ at oblique angles.\cite{Jin2019} Therefore, within a small angular window close to normal incidence, the quality factor $Q(k_x)$ of the LP can be determined from a relatively simple law:
	\begin{align}
	{Q(k_x)}^{-1}=Q_0^{-1} + \left(Q_\infty + \frac{\alpha}{k_x^2}\right)^{-1}\label{eq:Q},
	\end{align}
	in which $Q_0$ corresponds to the non-radiative losses originated from the excitonic component, $Q_\infty$ corresponds to the radiative losses of the photonic BIC at high oblique angles, and the coefficient $\alpha$ dictates the decreasing rate of the photonic BIC. Fig.~\ref{fig3}b depicts the angular dependence of the quality factor of the LPB that are extracted from the photoluminescence spectra. The experimental data are nicely fitted by the model \eqref{eq:Q}, with $Q_\infty=95$, $\alpha=5.9$\,$\mu$m$^2$ and $Q_0$>1000. Therefore, our pol-BIC exhibits a quasi-BIC nature, indeed, having infinite radiative quality factor inherited from the photonic component, but a finite non-radiative one (>1000) inherited from the excitonic component.
	
	Despite being quasi-BICs with finite quality factor, pol-BICs preserve the nature of singularities in momentum space. This is due to a complete decoupling from the radiative continuum. As a consequence, every polaritonic branch that hosts a pol-BIC  exhibits a polarization vortex around this singularity. In other words, pol-BICs fully inherit the topological behavior of their photonic components. To our knowledge, this fundamental property of pol-BICs has not been addressed in the literature, so far, either theoretically or experimentally. Here we investigate experimentally the polarization pattern of the LP branch in the vicinity of the pol-BIC$_{1}$. Figs.~\ref{fig4}a,b present the measured Stokes parameters $S_1$ (Fig.~\ref{fig4}a) and $S_2$ (Fig.~\ref{fig4}a). These results evidence clearly the existence of a polarization singularity at $\Gamma$ point.  We note that the third Stokes parameter, $S_3$ (not shown here), is measured to be negligible, indicating that the ellipticity of the farfield polarization is also negligible.  From the mapping of $S_1$ and $S_2$, we can easily extract the polarization orientation, $\phi$, given by the relation $\tan\{2\phi\} = S_2/S_1$. The latter is  reported in Fig.~\ref{fig4}c, showing that $\phi$ decreases by an angle $2\pi$ when the $\Gamma$ point is encircled. Therefore, the pol-BIC$_1$ is associated to a topological charge $q$, given by\cite{Zhen2014}: 
	
	\begin{equation}
	q = \frac{1}{2\pi} \oint_\mathcal{C} d\vec{k}.\nabla_{\vec{k}} \phi = -1.
	\end{equation}
	
	 Finally, we theoretically verify that such topological nature of pol-BICs is indeed originated from the purely photonic BIC components. We report in  Figs.~\ref{fig4}a',b',c' the theoretically determined Stokes parameters and the corresponding polarization orientation of the photonic BIC$_1$, calculated from the GME results of the passive photonic structure (see Fig.~\ref{fig2}a, and details in supplementary information). The striking similarity between these calculated patterns and the ones measured from pol-BIC$_1$ confirm that the photonic topological charge of this mode, $q=-1$, has been fully transferred to the polaritonic one.

	\section{\label{sec:level3} Conclusion}
	In conclusion, we have experimentally demonstrated the formation of polariton BICs at room temperature in a perovskite-based resonant metasurface. As hybrid eigenmodes, these excitations inherit the non-radiative losses from the excitonic component, but at the same time preserve the topological nature of the purely photonic BIC. Our results suggest that excitonic metasurfaces are a promising platform to reveal and exploit polariton effects (such as nonlinearity, or interaction with magnetic field) with the engineering of polarization singularities. Indeed, one may think of using the polariton nonlinearity to control the merging/splitting of topological charges\cite{Jin2019,Hwang2021}, or exploring the multi-bistable regimes of BIC physics\cite{BIC_nonlinear,BICnonlinear2}. Moreover, this would pave the way to explore new regimes of polariton condensation and quantum fluids of light, in which the condensate is totally decoupled from the radiative continuum and is purely non-radiative.
	
	\section{\label{sec:level4}Methods}
	\subsection{\label{sec:level5}Sample fabrication}
	The 2D structures were patterned using the laser interference lithography (LIL) technique. Substrates of 2~$\mu$m SiO$_2$/Si with size 1.5~x~1.5~cm were cleaned with acetone, ethanol, isopropanol then spincoated with photoresist MaN2403 at 5000~rpm in 30~sec then baked at 90$^o$C in 1~min. These substrates were then put inside the LIL setup, exposed under laser $\lambda$=266~nm with power 8~mW. Following this, we developed the samples in AZ- MIF 726 solution in 20~s. After LIL process, the patterns were formed on photoresist film and later transferred on SiO$_2$ using reactive ion etching (RIE) method. SiO$_2$ was etched by a gas combination of CHF$_3$/SF$_6$/CH$_4$ with respectively 16:6.2:2~sccm at pressure 20~mTorr with power set to 60~W. Followed by the fabrication of 2D patterns, PEPI 10\% in DMF was deposited on the patterns by spincoating at 3000~rpm in 30~sec then annealed at 95°C in 90~sec. Last, we encapsulated PEPI from oxygen and humidity by spin coating PMMA resist on top of the whole structure at 5000~rpm in 30~sec and baking at 95°C in 5~min. 
	
	\subsection{\label{sec:level6}Fourier spectroscopy}
	The sample is excited by a halogen lamp (for reflectivity and resonant scattering measurements) or by non-resonant laser at 400\,nm (for photoluminescence measurements) through a microscope objective ($\times$20, NA=0.42). The scattered/emitter signal is collected via the same microscope objective and then is analysed in the momentum space by imaging the backfocal plane of the microscope objective onto the camera sensor. For dispersion measurement, the signal is first projected onto the entrance a spectrometer of which the output is coupled to the sensor of a CCD camera. For the farfield imaging, the signal is spectrally filtered by a band pass filter within 544~nm - 556~nm, and then projected onto the sensor of a SCMOS camera. Further information of the experimental setup is detailed in the Supplemental Material.
	
	\subsection{\label{sec:level7}Polarization mapping}
	The mapping of Stokes parameters, $S_1(k_x,k_y)$, $S_2(k_x,k_y)$ and $S_3(k_x,k_y)$, is directly extracted from the photoluminescence farfield images, which are analysed in six different  polarizations, respectively $\Hpol$, $\Vpol$, $\Dpol$, $\Apol$, $\lcirclearrowright$, $\lcirclearrowleft$. Then, the ellipticity, $\chi$, is extracted from $S_3$: $\sin\{2\chi\} = S_3$; the orientation, $\phi$, is extracted from $S_1$ and $S_2$ as $\tan\{2\phi\} = S_2/S_1$. 
	
	\subsection{\label{sec:level8}Theoretical methods}
	The GME calculations have been performed by assuming an asymmetric planar waveguide with air (silica) as top (bottom) cladding, a core made of 200 nm thick PMMA (refractive index 1.49), 30 nm of unpatterned PEPI layer (index 2.41), and a square photonic lattice of 150 nm-thick circular PEPI pillars in a silica (index 1.48) backbone, lattice constant 330 nm, and pillar radius $r/a=0.2$. The Stokes parameters are then calculated from the corresponding Bloch modes as\cite{Yoda2020}
	$c_x(k_x,k_y)=\hat{x} \cdot \langle \vec{u}(k_x,k_y) \rangle $ and $c_y(k_x,k_y)=\hat{y} \cdot \langle \vec{u}(k_x,k_y) \rangle $, where the Bloch functions $\vec{u}$ are calculated within the metasurface core (specifically, at half thickness of the patterned silica backbone), and the average $\langle \vec{u} \rangle$ is performed on the direct lattice unit cell. From these, we calculate $S_1 = |c_x|^2-|c_y|^2$, $S_2 = 2\, \mathrm{Re}\{c_x^{\ast} c_y \}$, and $S_3 = 2\, \mathrm{Im}\{c_x^{\ast} c_y \}$. Finally, the Stokes parameters are normalized as $(S_1^2+S_2^2+S_3^2)^{1/2}=1$. 
	
	The RCWA simulations of angular-resolved reflectivity have been performed with the S\textsuperscript{4} package provided by the Fan Group at the Stanford Electrical Engineering Department\cite{Liu2012}. The dielectric function of PEPI, used in  RCWA simulations, is given by:
	\begin{align}
	\epsilon_{PEPI}(E)=n^2+\frac{A_X}{E_X^2-E^2-i\gamma_XE}
	\end{align}
	where $n=2.4$ is the refractive index of the passive structure, $A_X=0.85$\,eV$^2$ is the oscillator strength of PEPI exciton, $E_X=2.394$\,eV is its energy and $\gamma_X=30$\,meV is its linewidth~\cite{Dang2020}.
	
	We notice that the Rabi coupling in polaritonic systems with confined excitons can be expressed in terms of the bulk value as
	\begin{align}
	\hbar\omega_c\simeq \frac{\hbar\Omega}{2} = \hbar\omega_{\mathrm{bulk}}\sqrt{\frac{L_w}{L_{\mathrm{eff}}}}
	\end{align}
	in which $\hbar\omega_{\mathrm{bulk}}=\sqrt{A_X}/n$, $L_w$ is the thickness of the active layer (i.e., where the field is mostly intense), and $L_{\mathrm{eff}}$ is the effective extension length of the confined electric field. The latter quantities can be estimated from the field profile in  Fig.~2b, i.e., $L_w \sim 30$ nm, $L_{\mathrm{eff}} \sim 300$ nm. With these estimates we get $\hbar\Omega \simeq 240$ meV, in good agreement with the experimentally determined value.
	
	\begin{acknowledgments}
		
		The authors would like to thank the staff from the Nanolyon Technical Platform for helping and supporting in all nanofabrication processes. This work is partly supported by the French National Research Agency (ANR) under the project POPEYE (ANR-17-CE24-0020), project EMIPERO (ANR-18-CE24-0016) and the IDEXLYON from Université de Lyon, Scientific Breakthrough project TORE within the Programme Investissements d’Avenir (ANR-19-IDEX-0005). DS and DG  acknowledge financial support from the Italian Ministry of Research (MIUR) through the PRIN 2017 project ``Interacting photons in polariton circuits'' (INPhoPOL).
		
	\end{acknowledgments}
	
	\nocite{*}
	\bibliography{main}

\begin{thebibliography}{54}%
\makeatletter
\providecommand \@ifxundefined [1]{%
 \@ifx{#1\undefined}
}%
\providecommand \@ifnum [1]{%
 \ifnum #1\expandafter \@firstoftwo
 \else \expandafter \@secondoftwo
 \fi
}%
\providecommand \@ifx [1]{%
 \ifx #1\expandafter \@firstoftwo
 \else \expandafter \@secondoftwo
 \fi
}%
\providecommand \natexlab [1]{#1}%
\providecommand \enquote  [1]{``#1''}%
\providecommand \bibnamefont  [1]{#1}%
\providecommand \bibfnamefont [1]{#1}%
\providecommand \citenamefont [1]{#1}%
\providecommand \href@noop [0]{\@secondoftwo}%
\providecommand \href [0]{\begingroup \@sanitize@url \@href}%
\providecommand \@href[1]{\@@startlink{#1}\@@href}%
\providecommand \@@href[1]{\endgroup#1\@@endlink}%
\providecommand \@sanitize@url [0]{\catcode `\\12\catcode `\$12\catcode
  `\&12\catcode `\#12\catcode `\^12\catcode `\_12\catcode `\%12\relax}%
\providecommand \@@startlink[1]{}%
\providecommand \@@endlink[0]{}%
\providecommand \url  [0]{\begingroup\@sanitize@url \@url }%
\providecommand \@url [1]{\endgroup\@href {#1}{\urlprefix }}%
\providecommand \urlprefix  [0]{URL }%
\providecommand \Eprint [0]{\href }%
\providecommand \doibase [0]{http://dx.doi.org/}%
\providecommand \selectlanguage [0]{\@gobble}%
\providecommand \bibinfo  [0]{\@secondoftwo}%
\providecommand \bibfield  [0]{\@secondoftwo}%
\providecommand \translation [1]{[#1]}%
\providecommand \BibitemOpen [0]{}%
\providecommand \bibitemStop [0]{}%
\providecommand \bibitemNoStop [0]{.\EOS\space}%
\providecommand \EOS [0]{\spacefactor3000\relax}%
\providecommand \BibitemShut  [1]{\csname bibitem#1\endcsname}%
\let\auto@bib@innerbib\@empty
\bibitem [{\citenamefont {{Von Neumann}}\ and\ \citenamefont
  {Wigner}(1929)}]{VonNeumann1929}%
  \BibitemOpen
  \bibfield  {author} {\bibinfo {author} {\bibfnamefont {J.}~\bibnamefont {{Von
  Neumann}}}\ and\ \bibinfo {author} {\bibfnamefont {E.}~\bibnamefont
  {Wigner}},\ }\bibfield  {title} {\enquote {\bibinfo {title} {{{\"{U}}ber
  merkw{\"{u}}rdige diskrete Eigenwerte}},}\ }\href@noop {} {\bibfield
  {journal} {\bibinfo  {journal} {Z. Phys}\ }\textbf {\bibinfo {volume} {30}},\
  \bibinfo {pages} {465--497} (\bibinfo {year} {1929})}\BibitemShut {NoStop}%
\bibitem [{\citenamefont {Hsu}\ \emph {et~al.}(2016)\citenamefont {Hsu},
  \citenamefont {Zhen}, \citenamefont {Stone}, \citenamefont {Joannopoulos},\
  and\ \citenamefont {Solja{\v{c}}i{\'{c}}}}]{Hsu2016}%
  \BibitemOpen
  \bibfield  {author} {\bibinfo {author} {\bibfnamefont {C.~W.}\ \bibnamefont
  {Hsu}}, \bibinfo {author} {\bibfnamefont {B.}~\bibnamefont {Zhen}}, \bibinfo
  {author} {\bibfnamefont {A.~D.}\ \bibnamefont {Stone}}, \bibinfo {author}
  {\bibfnamefont {J.~D.}\ \bibnamefont {Joannopoulos}}, \ and\ \bibinfo
  {author} {\bibfnamefont {M.}~\bibnamefont {Solja{\v{c}}i{\'{c}}}},\
  }\bibfield  {title} {\enquote {\bibinfo {title} {{Bound states in the
  continuum}},}\ }\href {\doibase 10.1038/natrevmats.2016.48} {\bibfield
  {journal} {\bibinfo  {journal} {Nature Reviews Materials}\ }\textbf {\bibinfo
  {volume} {1}},\ \bibinfo {pages} {16048} (\bibinfo {year}
  {2016})}\BibitemShut {NoStop}%
\bibitem [{\citenamefont {Azzam}\ and\ \citenamefont
  {Kildishev}(2021)}]{Azzam2021}%
  \BibitemOpen
  \bibfield  {author} {\bibinfo {author} {\bibfnamefont {S.~I.}\ \bibnamefont
  {Azzam}}\ and\ \bibinfo {author} {\bibfnamefont {A.~V.}\ \bibnamefont
  {Kildishev}},\ }\bibfield  {title} {\enquote {\bibinfo {title} {Photonic
  bound states in the continuum: From basics to applications},}\ }\href
  {\doibase https://doi.org/10.1002/adom.202001469} {\bibfield  {journal}
  {\bibinfo  {journal} {Advanced Optical Materials}\ }\textbf {\bibinfo
  {volume} {9}},\ \bibinfo {pages} {2001469} (\bibinfo {year} {2021})},\
  \Eprint
  {http://arxiv.org/abs/https://onlinelibrary.wiley.com/doi/pdf/10.1002/adom.202001469}
  {https://onlinelibrary.wiley.com/doi/pdf/10.1002/adom.202001469} \BibitemShut
  {NoStop}%
\bibitem [{\citenamefont {Kodigala}\ \emph {et~al.}(2017)\citenamefont
  {Kodigala}, \citenamefont {Lepetit}, \citenamefont {Gu}, \citenamefont
  {Bahari}, \citenamefont {Fainman},\ and\ \citenamefont
  {Kant{\'{e}}}}]{Kodigala2017}%
  \BibitemOpen
  \bibfield  {author} {\bibinfo {author} {\bibfnamefont {A.}~\bibnamefont
  {Kodigala}}, \bibinfo {author} {\bibfnamefont {T.}~\bibnamefont {Lepetit}},
  \bibinfo {author} {\bibfnamefont {Q.}~\bibnamefont {Gu}}, \bibinfo {author}
  {\bibfnamefont {B.}~\bibnamefont {Bahari}}, \bibinfo {author} {\bibfnamefont
  {Y.}~\bibnamefont {Fainman}}, \ and\ \bibinfo {author} {\bibfnamefont
  {B.}~\bibnamefont {Kant{\'{e}}}},\ }\bibfield  {title} {\enquote {\bibinfo
  {title} {{Lasing action from photonic bound states in continuum}},}\ }\href
  {\doibase 10.1038/nature20799} {\bibfield  {journal} {\bibinfo  {journal}
  {Nature}\ }\textbf {\bibinfo {volume} {541}},\ \bibinfo {pages} {196--199}
  (\bibinfo {year} {2017})}\BibitemShut {NoStop}%
\bibitem [{\citenamefont {Jin}\ \emph {et~al.}(2019)\citenamefont {Jin},
  \citenamefont {Yin}, \citenamefont {Ni}, \citenamefont
  {Solja{\v{c}}i{\'{c}}}, \citenamefont {Zhen},\ and\ \citenamefont
  {Peng}}]{Jin2019}%
  \BibitemOpen
  \bibfield  {author} {\bibinfo {author} {\bibfnamefont {J.}~\bibnamefont
  {Jin}}, \bibinfo {author} {\bibfnamefont {X.}~\bibnamefont {Yin}}, \bibinfo
  {author} {\bibfnamefont {L.}~\bibnamefont {Ni}}, \bibinfo {author}
  {\bibfnamefont {M.}~\bibnamefont {Solja{\v{c}}i{\'{c}}}}, \bibinfo {author}
  {\bibfnamefont {B.}~\bibnamefont {Zhen}}, \ and\ \bibinfo {author}
  {\bibfnamefont {C.}~\bibnamefont {Peng}},\ }\bibfield  {title} {\enquote
  {\bibinfo {title} {{Topologically enabled ultrahigh-Q guided resonances
  robust to out-of-plane scattering}},}\ }\href {\doibase
  10.1038/s41586-019-1664-7} {\bibfield  {journal} {\bibinfo  {journal}
  {Nature}\ }\textbf {\bibinfo {volume} {574}},\ \bibinfo {pages} {501--504}
  (\bibinfo {year} {2019})}\BibitemShut {NoStop}%
\bibitem [{\citenamefont {Wu}\ \emph {et~al.}(2020)\citenamefont {Wu},
  \citenamefont {Ha}, \citenamefont {Shendre}, \citenamefont {Durmusoglu},
  \citenamefont {Koh}, \citenamefont {Abujetas}, \citenamefont
  {S{\'{a}}nchez-Gil}, \citenamefont {Paniagua-Dom{\'{i}}nguez}, \citenamefont
  {Demir},\ and\ \citenamefont {Kuznetsov}}]{Wu2020}%
  \BibitemOpen
  \bibfield  {author} {\bibinfo {author} {\bibfnamefont {M.}~\bibnamefont
  {Wu}}, \bibinfo {author} {\bibfnamefont {S.~T.}\ \bibnamefont {Ha}}, \bibinfo
  {author} {\bibfnamefont {S.}~\bibnamefont {Shendre}}, \bibinfo {author}
  {\bibfnamefont {E.~G.}\ \bibnamefont {Durmusoglu}}, \bibinfo {author}
  {\bibfnamefont {W.-K.}\ \bibnamefont {Koh}}, \bibinfo {author} {\bibfnamefont
  {D.~R.}\ \bibnamefont {Abujetas}}, \bibinfo {author} {\bibfnamefont {J.~A.}\
  \bibnamefont {S{\'{a}}nchez-Gil}}, \bibinfo {author} {\bibfnamefont
  {R.}~\bibnamefont {Paniagua-Dom{\'{i}}nguez}}, \bibinfo {author}
  {\bibfnamefont {H.~V.}\ \bibnamefont {Demir}}, \ and\ \bibinfo {author}
  {\bibfnamefont {A.~I.}\ \bibnamefont {Kuznetsov}},\ }\bibfield  {title}
  {\enquote {\bibinfo {title} {{Room-Temperature Lasing in Colloidal
  Nanoplatelets via Mie-Resonant Bound States in the Continuum}},}\ }\href
  {\doibase 10.1021/acs.nanolett.0c01975} {\bibfield  {journal} {\bibinfo
  {journal} {Nano Letters}\ }\textbf {\bibinfo {volume} {20}},\ \bibinfo
  {pages} {6005--6011} (\bibinfo {year} {2020})}\BibitemShut {NoStop}%
\bibitem [{\citenamefont {Huang}\ \emph {et~al.}(2020)\citenamefont {Huang},
  \citenamefont {Zhang}, \citenamefont {Xiao}, \citenamefont {Wang},
  \citenamefont {Fan}, \citenamefont {Liu}, \citenamefont {Zhang},
  \citenamefont {Qu}, \citenamefont {Ji}, \citenamefont {Han}, \citenamefont
  {Ge}, \citenamefont {Kivshar},\ and\ \citenamefont {Song}}]{BICSwitch}%
  \BibitemOpen
  \bibfield  {author} {\bibinfo {author} {\bibfnamefont {C.}~\bibnamefont
  {Huang}}, \bibinfo {author} {\bibfnamefont {C.}~\bibnamefont {Zhang}},
  \bibinfo {author} {\bibfnamefont {S.}~\bibnamefont {Xiao}}, \bibinfo {author}
  {\bibfnamefont {Y.}~\bibnamefont {Wang}}, \bibinfo {author} {\bibfnamefont
  {Y.}~\bibnamefont {Fan}}, \bibinfo {author} {\bibfnamefont {Y.}~\bibnamefont
  {Liu}}, \bibinfo {author} {\bibfnamefont {N.}~\bibnamefont {Zhang}}, \bibinfo
  {author} {\bibfnamefont {G.}~\bibnamefont {Qu}}, \bibinfo {author}
  {\bibfnamefont {H.}~\bibnamefont {Ji}}, \bibinfo {author} {\bibfnamefont
  {J.}~\bibnamefont {Han}}, \bibinfo {author} {\bibfnamefont {L.}~\bibnamefont
  {Ge}}, \bibinfo {author} {\bibfnamefont {Y.}~\bibnamefont {Kivshar}}, \ and\
  \bibinfo {author} {\bibfnamefont {Q.}~\bibnamefont {Song}},\ }\bibfield
  {title} {\enquote {\bibinfo {title} {Ultrafast control of vortex
  microlasers},}\ }\href {\doibase 10.1126/science.aba4597} {\bibfield
  {journal} {\bibinfo  {journal} {Science}\ }\textbf {\bibinfo {volume}
  {367}},\ \bibinfo {pages} {1018--1021} (\bibinfo {year} {2020})}\BibitemShut
  {NoStop}%
\bibitem [{\citenamefont {Hwang}\ \emph {et~al.}(2021)\citenamefont {Hwang},
  \citenamefont {Lee}, \citenamefont {Kim}, \citenamefont {Jeong},
  \citenamefont {Kwon}, \citenamefont {Koshelev}, \citenamefont {Kivshar},\
  and\ \citenamefont {Park}}]{Hwang2021}%
  \BibitemOpen
  \bibfield  {author} {\bibinfo {author} {\bibfnamefont {M.-S.}\ \bibnamefont
  {Hwang}}, \bibinfo {author} {\bibfnamefont {H.-C.}\ \bibnamefont {Lee}},
  \bibinfo {author} {\bibfnamefont {K.-H.}\ \bibnamefont {Kim}}, \bibinfo
  {author} {\bibfnamefont {K.-Y.}\ \bibnamefont {Jeong}}, \bibinfo {author}
  {\bibfnamefont {S.-H.}\ \bibnamefont {Kwon}}, \bibinfo {author}
  {\bibfnamefont {K.}~\bibnamefont {Koshelev}}, \bibinfo {author}
  {\bibfnamefont {Y.}~\bibnamefont {Kivshar}}, \ and\ \bibinfo {author}
  {\bibfnamefont {H.-G.}\ \bibnamefont {Park}},\ }\bibfield  {title} {\enquote
  {\bibinfo {title} {{Ultralow-threshold laser using super-bound states in the
  continuum}},}\ }\href {\doibase 10.1038/s41467-021-24502-0} {\bibfield
  {journal} {\bibinfo  {journal} {Nature Communications}\ }\textbf {\bibinfo
  {volume} {12}},\ \bibinfo {pages} {4135} (\bibinfo {year}
  {2021})}\BibitemShut {NoStop}%
\bibitem [{\citenamefont {Romano}\ \emph {et~al.}(2018)\citenamefont {Romano},
  \citenamefont {Zito}, \citenamefont {Torino}, \citenamefont {Calafiore},
  \citenamefont {Penzo}, \citenamefont {Coppola}, \citenamefont {Cabrini},
  \citenamefont {Rendina},\ and\ \citenamefont {Mocella}}]{Romano:18}%
  \BibitemOpen
  \bibfield  {author} {\bibinfo {author} {\bibfnamefont {S.}~\bibnamefont
  {Romano}}, \bibinfo {author} {\bibfnamefont {G.}~\bibnamefont {Zito}},
  \bibinfo {author} {\bibfnamefont {S.}~\bibnamefont {Torino}}, \bibinfo
  {author} {\bibfnamefont {G.}~\bibnamefont {Calafiore}}, \bibinfo {author}
  {\bibfnamefont {E.}~\bibnamefont {Penzo}}, \bibinfo {author} {\bibfnamefont
  {G.}~\bibnamefont {Coppola}}, \bibinfo {author} {\bibfnamefont
  {S.}~\bibnamefont {Cabrini}}, \bibinfo {author} {\bibfnamefont
  {I.}~\bibnamefont {Rendina}}, \ and\ \bibinfo {author} {\bibfnamefont
  {V.}~\bibnamefont {Mocella}},\ }\bibfield  {title} {\enquote {\bibinfo
  {title} {Label-free sensing of ultralow-weight molecules with all-dielectric
  metasurfaces supporting bound states in the continuum},}\ }\href {\doibase
  10.1364/PRJ.6.000726} {\bibfield  {journal} {\bibinfo  {journal} {Photon.
  Res.}\ }\textbf {\bibinfo {volume} {6}},\ \bibinfo {pages} {726--733}
  (\bibinfo {year} {2018})}\BibitemShut {NoStop}%
\bibitem [{\citenamefont {Krasikov}, \citenamefont {Bogdanov},\ and\
  \citenamefont {Iorsh}(2018)}]{BIC_nonlinear}%
  \BibitemOpen
  \bibfield  {author} {\bibinfo {author} {\bibfnamefont {S.~D.}\ \bibnamefont
  {Krasikov}}, \bibinfo {author} {\bibfnamefont {A.~A.}\ \bibnamefont
  {Bogdanov}}, \ and\ \bibinfo {author} {\bibfnamefont {I.~V.}\ \bibnamefont
  {Iorsh}},\ }\bibfield  {title} {\enquote {\bibinfo {title} {Nonlinear bound
  states in the continuum of a one-dimensional photonic crystal slab},}\ }\href
  {\doibase 10.1103/PhysRevB.97.224309} {\bibfield  {journal} {\bibinfo
  {journal} {Phys. Rev. B}\ }\textbf {\bibinfo {volume} {97}},\ \bibinfo
  {pages} {224309} (\bibinfo {year} {2018})}\BibitemShut {NoStop}%
\bibitem [{\citenamefont {Minkov}, \citenamefont {Gerace},\ and\ \citenamefont
  {Fan}(2019)}]{Minkov:19}%
  \BibitemOpen
  \bibfield  {author} {\bibinfo {author} {\bibfnamefont {M.}~\bibnamefont
  {Minkov}}, \bibinfo {author} {\bibfnamefont {D.}~\bibnamefont {Gerace}}, \
  and\ \bibinfo {author} {\bibfnamefont {S.}~\bibnamefont {Fan}},\ }\bibfield
  {title} {\enquote {\bibinfo {title} {Doubly resonant $\chi^{(2)}$ nonlinear
  photonic crystal cavity based on a bound state in the continuum},}\ }\href
  {\doibase 10.1364/OPTICA.6.001039} {\bibfield  {journal} {\bibinfo  {journal}
  {Optica}\ }\textbf {\bibinfo {volume} {6}},\ \bibinfo {pages} {1039--1045}
  (\bibinfo {year} {2019})}\BibitemShut {NoStop}%
\bibitem [{\citenamefont {Zakharov}\ and\ \citenamefont
  {Poddubny}(2020)}]{BICnonlinear2}%
  \BibitemOpen
  \bibfield  {author} {\bibinfo {author} {\bibfnamefont {V.~A.}\ \bibnamefont
  {Zakharov}}\ and\ \bibinfo {author} {\bibfnamefont {A.~N.}\ \bibnamefont
  {Poddubny}},\ }\bibfield  {title} {\enquote {\bibinfo {title} {Transverse
  magneto-optical kerr effect enhanced at the bound states in the continuum},}\
  }\href {\doibase 10.1103/PhysRevA.101.043848} {\bibfield  {journal} {\bibinfo
   {journal} {Phys. Rev. A}\ }\textbf {\bibinfo {volume} {101}},\ \bibinfo
  {pages} {043848} (\bibinfo {year} {2020})}\BibitemShut {NoStop}%
\bibitem [{\citenamefont {Wang}\ \emph
  {et~al.}(2020{\natexlab{a}})\citenamefont {Wang}, \citenamefont {Clementi},
  \citenamefont {Minkov}, \citenamefont {Barone}, \citenamefont {Carlin},
  \citenamefont {Grandjean}, \citenamefont {Gerace}, \citenamefont {Fan},
  \citenamefont {Galli},\ and\ \citenamefont {Houdr\'{e}}}]{Houdre2020}%
  \BibitemOpen
  \bibfield  {author} {\bibinfo {author} {\bibfnamefont {J.}~\bibnamefont
  {Wang}}, \bibinfo {author} {\bibfnamefont {M.}~\bibnamefont {Clementi}},
  \bibinfo {author} {\bibfnamefont {M.}~\bibnamefont {Minkov}}, \bibinfo
  {author} {\bibfnamefont {A.}~\bibnamefont {Barone}}, \bibinfo {author}
  {\bibfnamefont {J.-F.}\ \bibnamefont {Carlin}}, \bibinfo {author}
  {\bibfnamefont {N.}~\bibnamefont {Grandjean}}, \bibinfo {author}
  {\bibfnamefont {D.}~\bibnamefont {Gerace}}, \bibinfo {author} {\bibfnamefont
  {S.}~\bibnamefont {Fan}}, \bibinfo {author} {\bibfnamefont {M.}~\bibnamefont
  {Galli}}, \ and\ \bibinfo {author} {\bibfnamefont {R.}~\bibnamefont
  {Houdr\'{e}}},\ }\bibfield  {title} {\enquote {\bibinfo {title} {Doubly
  resonant second-harmonic generation of a vortex beam from a bound state in
  the continuum},}\ }\href {\doibase 10.1364/OPTICA.396408} {\bibfield
  {journal} {\bibinfo  {journal} {Optica}\ }\textbf {\bibinfo {volume} {7}},\
  \bibinfo {pages} {1126--1132} (\bibinfo {year}
  {2020}{\natexlab{a}})}\BibitemShut {NoStop}%
\bibitem [{\citenamefont {Zhen}\ \emph {et~al.}(2014)\citenamefont {Zhen},
  \citenamefont {Hsu}, \citenamefont {Lu}, \citenamefont {Stone},\ and\
  \citenamefont {Solja{\v{c}}i{\'{c}}}}]{Zhen2014}%
  \BibitemOpen
  \bibfield  {author} {\bibinfo {author} {\bibfnamefont {B.}~\bibnamefont
  {Zhen}}, \bibinfo {author} {\bibfnamefont {C.~W.}\ \bibnamefont {Hsu}},
  \bibinfo {author} {\bibfnamefont {L.}~\bibnamefont {Lu}}, \bibinfo {author}
  {\bibfnamefont {A.~D.}\ \bibnamefont {Stone}}, \ and\ \bibinfo {author}
  {\bibfnamefont {M.}~\bibnamefont {Solja{\v{c}}i{\'{c}}}},\ }\bibfield
  {title} {\enquote {\bibinfo {title} {{Topological Nature of Optical Bound
  States in the Continuum}},}\ }\href {\doibase 10.1103/PhysRevLett.113.257401}
  {\bibfield  {journal} {\bibinfo  {journal} {Physical Review Letters}\
  }\textbf {\bibinfo {volume} {113}},\ \bibinfo {pages} {1--5} (\bibinfo {year}
  {2014})}\BibitemShut {NoStop}%
\bibitem [{\citenamefont {Doeleman}\ \emph {et~al.}(2018)\citenamefont
  {Doeleman}, \citenamefont {Monticone}, \citenamefont {den Hollander},
  \citenamefont {Al{\`{u}}},\ and\ \citenamefont {Koenderink}}]{Doeleman2018}%
  \BibitemOpen
  \bibfield  {author} {\bibinfo {author} {\bibfnamefont {H.~M.}\ \bibnamefont
  {Doeleman}}, \bibinfo {author} {\bibfnamefont {F.}~\bibnamefont {Monticone}},
  \bibinfo {author} {\bibfnamefont {W.}~\bibnamefont {den Hollander}}, \bibinfo
  {author} {\bibfnamefont {A.}~\bibnamefont {Al{\`{u}}}}, \ and\ \bibinfo
  {author} {\bibfnamefont {A.~F.}\ \bibnamefont {Koenderink}},\ }\bibfield
  {title} {\enquote {\bibinfo {title} {{Experimental observation of a
  polarization vortex at an optical bound state in the continuum}},}\ }\href
  {\doibase 10.1038/s41566-018-0177-5} {\bibfield  {journal} {\bibinfo
  {journal} {Nature Photonics}\ }\textbf {\bibinfo {volume} {12}},\ \bibinfo
  {pages} {397--401} (\bibinfo {year} {2018})}\BibitemShut {NoStop}%
\bibitem [{\citenamefont {Zhang}\ \emph {et~al.}(2018)\citenamefont {Zhang},
  \citenamefont {Chen}, \citenamefont {Liu}, \citenamefont {Hsu}, \citenamefont
  {Wang}, \citenamefont {Guan}, \citenamefont {Liu}, \citenamefont {Shi},
  \citenamefont {Lu},\ and\ \citenamefont {Zi}}]{Zhang2018}%
  \BibitemOpen
  \bibfield  {author} {\bibinfo {author} {\bibfnamefont {Y.}~\bibnamefont
  {Zhang}}, \bibinfo {author} {\bibfnamefont {A.}~\bibnamefont {Chen}},
  \bibinfo {author} {\bibfnamefont {W.}~\bibnamefont {Liu}}, \bibinfo {author}
  {\bibfnamefont {C.~W.}\ \bibnamefont {Hsu}}, \bibinfo {author} {\bibfnamefont
  {B.}~\bibnamefont {Wang}}, \bibinfo {author} {\bibfnamefont {F.}~\bibnamefont
  {Guan}}, \bibinfo {author} {\bibfnamefont {X.}~\bibnamefont {Liu}}, \bibinfo
  {author} {\bibfnamefont {L.}~\bibnamefont {Shi}}, \bibinfo {author}
  {\bibfnamefont {L.}~\bibnamefont {Lu}}, \ and\ \bibinfo {author}
  {\bibfnamefont {J.}~\bibnamefont {Zi}},\ }\bibfield  {title} {\enquote
  {\bibinfo {title} {Observation of polarization vortices in momentum space},}\
  }\href {\doibase 10.1103/PhysRevLett.120.186103} {\bibfield  {journal}
  {\bibinfo  {journal} {Phys. Rev. Lett.}\ }\textbf {\bibinfo {volume} {120}},\
  \bibinfo {pages} {186103} (\bibinfo {year} {2018})}\BibitemShut {NoStop}%
\bibitem [{\citenamefont {Wang}\ \emph
  {et~al.}(2020{\natexlab{b}})\citenamefont {Wang}, \citenamefont {Liu},
  \citenamefont {Zhao}, \citenamefont {Wang}, \citenamefont {Zhang},
  \citenamefont {Chen}, \citenamefont {Guan}, \citenamefont {Liu},
  \citenamefont {Shi},\ and\ \citenamefont {Zi}}]{Wang2020}%
  \BibitemOpen
  \bibfield  {author} {\bibinfo {author} {\bibfnamefont {B.}~\bibnamefont
  {Wang}}, \bibinfo {author} {\bibfnamefont {W.}~\bibnamefont {Liu}}, \bibinfo
  {author} {\bibfnamefont {M.}~\bibnamefont {Zhao}}, \bibinfo {author}
  {\bibfnamefont {J.}~\bibnamefont {Wang}}, \bibinfo {author} {\bibfnamefont
  {Y.}~\bibnamefont {Zhang}}, \bibinfo {author} {\bibfnamefont
  {A.}~\bibnamefont {Chen}}, \bibinfo {author} {\bibfnamefont {F.}~\bibnamefont
  {Guan}}, \bibinfo {author} {\bibfnamefont {X.}~\bibnamefont {Liu}}, \bibinfo
  {author} {\bibfnamefont {L.}~\bibnamefont {Shi}}, \ and\ \bibinfo {author}
  {\bibfnamefont {J.}~\bibnamefont {Zi}},\ }\bibfield  {title} {\enquote
  {\bibinfo {title} {{Generating optical vortex beams by momentum-space
  polarization vortices centred at bound states in the continuum}},}\ }\href
  {\doibase 10.1038/s41566-020-0658-1} {\bibfield  {journal} {\bibinfo
  {journal} {Nature Photonics}\ }\textbf {\bibinfo {volume} {14}},\ \bibinfo
  {pages} {623--628} (\bibinfo {year} {2020}{\natexlab{b}})}\BibitemShut
  {NoStop}%
\bibitem [{\citenamefont {Yin}\ \emph {et~al.}(2019)\citenamefont {Yin},
  \citenamefont {Jin}, \citenamefont {Soljačić}, \citenamefont {Peng},\ and\
  \citenamefont {Zhen}}]{yin2019}%
  \BibitemOpen
  \bibfield  {author} {\bibinfo {author} {\bibfnamefont {X.}~\bibnamefont
  {Yin}}, \bibinfo {author} {\bibfnamefont {J.}~\bibnamefont {Jin}}, \bibinfo
  {author} {\bibfnamefont {M.}~\bibnamefont {Soljačić}}, \bibinfo {author}
  {\bibfnamefont {C.}~\bibnamefont {Peng}}, \ and\ \bibinfo {author}
  {\bibfnamefont {B.}~\bibnamefont {Zhen}},\ }\href@noop {} {\enquote {\bibinfo
  {title} {Observation of unidirectional bound states in the continuum enabled
  by topological defects},}\ } (\bibinfo {year} {2019}),\ \Eprint
  {http://arxiv.org/abs/1904.11464} {arXiv:1904.11464 [physics.optics]}
  \BibitemShut {NoStop}%
\bibitem [{\citenamefont {Ye}, \citenamefont {Gao},\ and\ \citenamefont
  {Liu}(2020)}]{Ye2020}%
  \BibitemOpen
  \bibfield  {author} {\bibinfo {author} {\bibfnamefont {W.}~\bibnamefont
  {Ye}}, \bibinfo {author} {\bibfnamefont {Y.}~\bibnamefont {Gao}}, \ and\
  \bibinfo {author} {\bibfnamefont {J.}~\bibnamefont {Liu}},\ }\bibfield
  {title} {\enquote {\bibinfo {title} {Singular points of polarizations in the
  momentum space of photonic crystal slabs},}\ }\href {\doibase
  10.1103/PhysRevLett.124.153904} {\bibfield  {journal} {\bibinfo  {journal}
  {Phys. Rev. Lett.}\ }\textbf {\bibinfo {volume} {124}},\ \bibinfo {pages}
  {153904} (\bibinfo {year} {2020})}\BibitemShut {NoStop}%
\bibitem [{\citenamefont {Yoda}\ and\ \citenamefont {Notomi}(2020)}]{Yoda2020}%
  \BibitemOpen
  \bibfield  {author} {\bibinfo {author} {\bibfnamefont {T.}~\bibnamefont
  {Yoda}}\ and\ \bibinfo {author} {\bibfnamefont {M.}~\bibnamefont {Notomi}},\
  }\bibfield  {title} {\enquote {\bibinfo {title} {Generation and annihilation
  of topologically protected bound states in the continuum and circularly
  polarized states by symmetry breaking},}\ }\href {\doibase
  10.1103/PhysRevLett.125.053902} {\bibfield  {journal} {\bibinfo  {journal}
  {Phys. Rev. Lett.}\ }\textbf {\bibinfo {volume} {125}},\ \bibinfo {pages}
  {053902} (\bibinfo {year} {2020})}\BibitemShut {NoStop}%
\bibitem [{\citenamefont {Weisbuch}\ \emph {et~al.}(1992)\citenamefont
  {Weisbuch}, \citenamefont {Nishioka}, \citenamefont {Ishikawa},\ and\
  \citenamefont {Arakawa}}]{Claude1994}%
  \BibitemOpen
  \bibfield  {author} {\bibinfo {author} {\bibfnamefont {C.}~\bibnamefont
  {Weisbuch}}, \bibinfo {author} {\bibfnamefont {M.}~\bibnamefont {Nishioka}},
  \bibinfo {author} {\bibfnamefont {A.}~\bibnamefont {Ishikawa}}, \ and\
  \bibinfo {author} {\bibfnamefont {Y.}~\bibnamefont {Arakawa}},\ }\bibfield
  {title} {\enquote {\bibinfo {title} {Observation of the coupled
  exciton-photon mode splitting in a semiconductor quantum microcavity},}\
  }\href {\doibase 10.1103/PhysRevLett.69.3314} {\bibfield  {journal} {\bibinfo
   {journal} {Phys. Rev. Lett.}\ }\textbf {\bibinfo {volume} {69}},\ \bibinfo
  {pages} {3314--3317} (\bibinfo {year} {1992})}\BibitemShut {NoStop}%
\bibitem [{\citenamefont {Kasprzak}\ \emph {et~al.}(2006)\citenamefont
  {Kasprzak}, \citenamefont {Richard}, \citenamefont {Kundermann},
  \citenamefont {Baas}, \citenamefont {Jeambrun}, \citenamefont {Keeling},
  \citenamefont {Marchetti}, \citenamefont {Szyma{\'{n}}ska}, \citenamefont
  {Andr{\'{e}}}, \citenamefont {Staehli}, \citenamefont {Savona}, \citenamefont
  {Littlewood}, \citenamefont {Deveaud},\ and\ \citenamefont
  {Dang}}]{Kasprzak2006}%
  \BibitemOpen
  \bibfield  {author} {\bibinfo {author} {\bibfnamefont {J.}~\bibnamefont
  {Kasprzak}}, \bibinfo {author} {\bibfnamefont {M.}~\bibnamefont {Richard}},
  \bibinfo {author} {\bibfnamefont {S.}~\bibnamefont {Kundermann}}, \bibinfo
  {author} {\bibfnamefont {A.}~\bibnamefont {Baas}}, \bibinfo {author}
  {\bibfnamefont {P.}~\bibnamefont {Jeambrun}}, \bibinfo {author}
  {\bibfnamefont {J.~M.~J.}\ \bibnamefont {Keeling}}, \bibinfo {author}
  {\bibfnamefont {F.~M.}\ \bibnamefont {Marchetti}}, \bibinfo {author}
  {\bibfnamefont {M.~H.}\ \bibnamefont {Szyma{\'{n}}ska}}, \bibinfo {author}
  {\bibfnamefont {R.}~\bibnamefont {Andr{\'{e}}}}, \bibinfo {author}
  {\bibfnamefont {J.~L.}\ \bibnamefont {Staehli}}, \bibinfo {author}
  {\bibfnamefont {V.}~\bibnamefont {Savona}}, \bibinfo {author} {\bibfnamefont
  {P.~B.}\ \bibnamefont {Littlewood}}, \bibinfo {author} {\bibfnamefont
  {B.}~\bibnamefont {Deveaud}}, \ and\ \bibinfo {author} {\bibfnamefont
  {L.~S.}\ \bibnamefont {Dang}},\ }\bibfield  {title} {\enquote {\bibinfo
  {title} {{Bose-Einstein condensation of exciton polaritons.}}}\ }\href
  {\doibase 10.1038/nature05131} {\bibfield  {journal} {\bibinfo  {journal}
  {Nature}\ }\textbf {\bibinfo {volume} {443}},\ \bibinfo {pages} {409--414}
  (\bibinfo {year} {2006})}\BibitemShut {NoStop}%
\bibitem [{\citenamefont {Carusotto}\ and\ \citenamefont
  {Ciuti}(2013)}]{QuantumFLuidsOfLight}%
  \BibitemOpen
  \bibfield  {author} {\bibinfo {author} {\bibfnamefont {I.}~\bibnamefont
  {Carusotto}}\ and\ \bibinfo {author} {\bibfnamefont {C.}~\bibnamefont
  {Ciuti}},\ }\bibfield  {title} {\enquote {\bibinfo {title} {Quantum fluids of
  light},}\ }\href {\doibase 10.1103/RevModPhys.85.299} {\bibfield  {journal}
  {\bibinfo  {journal} {Rev. Mod. Phys.}\ }\textbf {\bibinfo {volume} {85}},\
  \bibinfo {pages} {299--366} (\bibinfo {year} {2013})}\BibitemShut {NoStop}%
\bibitem [{\citenamefont {Christopoulos}\ \emph {et~al.}(2007)\citenamefont
  {Christopoulos}, \citenamefont {{Von H{\"{o}}gersthal}}, \citenamefont
  {Grundy}, \citenamefont {Lagoudakis}, \citenamefont {Kavokin}, \citenamefont
  {Baumberg}, \citenamefont {Christmann}, \citenamefont {Butt{\'{e}}},
  \citenamefont {Feltin}, \citenamefont {Carlin},\ and\ \citenamefont
  {Grandjean}}]{Christopoulos2007}%
  \BibitemOpen
  \bibfield  {author} {\bibinfo {author} {\bibfnamefont {S.}~\bibnamefont
  {Christopoulos}}, \bibinfo {author} {\bibfnamefont {G.~B.~H.}\ \bibnamefont
  {{Von H{\"{o}}gersthal}}}, \bibinfo {author} {\bibfnamefont {a.~J.~D.}\
  \bibnamefont {Grundy}}, \bibinfo {author} {\bibfnamefont {P.~G.}\
  \bibnamefont {Lagoudakis}}, \bibinfo {author} {\bibfnamefont {a.~V.}\
  \bibnamefont {Kavokin}}, \bibinfo {author} {\bibfnamefont {J.~J.}\
  \bibnamefont {Baumberg}}, \bibinfo {author} {\bibfnamefont {G.}~\bibnamefont
  {Christmann}}, \bibinfo {author} {\bibfnamefont {R.}~\bibnamefont
  {Butt{\'{e}}}}, \bibinfo {author} {\bibfnamefont {E.}~\bibnamefont {Feltin}},
  \bibinfo {author} {\bibfnamefont {J.~F.}\ \bibnamefont {Carlin}}, \ and\
  \bibinfo {author} {\bibfnamefont {N.}~\bibnamefont {Grandjean}},\ }\bibfield
  {title} {\enquote {\bibinfo {title} {{Room-temperature polariton lasing in
  semiconductor microcavities}},}\ }\href {\doibase
  10.1103/PhysRevLett.98.126405} {\bibfield  {journal} {\bibinfo  {journal}
  {Physical Review Letters}\ }\textbf {\bibinfo {volume} {98}},\ \bibinfo
  {pages} {1--4} (\bibinfo {year} {2007})}\BibitemShut {NoStop}%
\bibitem [{\citenamefont {Daskalakis}\ \emph {et~al.}(2013)\citenamefont
  {Daskalakis}, \citenamefont {Eldridge}, \citenamefont {Christmann},
  \citenamefont {Trichas}, \citenamefont {Murray}, \citenamefont {Iliopoulos},
  \citenamefont {Monroy}, \citenamefont {Pelekanos}, \citenamefont {Baumberg},\
  and\ \citenamefont {Savvidis}}]{Daskalakis2013}%
  \BibitemOpen
  \bibfield  {author} {\bibinfo {author} {\bibfnamefont {K.~S.}\ \bibnamefont
  {Daskalakis}}, \bibinfo {author} {\bibfnamefont {P.~S.}\ \bibnamefont
  {Eldridge}}, \bibinfo {author} {\bibfnamefont {G.}~\bibnamefont
  {Christmann}}, \bibinfo {author} {\bibfnamefont {E.}~\bibnamefont {Trichas}},
  \bibinfo {author} {\bibfnamefont {R.}~\bibnamefont {Murray}}, \bibinfo
  {author} {\bibfnamefont {E.}~\bibnamefont {Iliopoulos}}, \bibinfo {author}
  {\bibfnamefont {E.}~\bibnamefont {Monroy}}, \bibinfo {author} {\bibfnamefont
  {N.~T.}\ \bibnamefont {Pelekanos}}, \bibinfo {author} {\bibfnamefont {J.~J.}\
  \bibnamefont {Baumberg}}, \ and\ \bibinfo {author} {\bibfnamefont {P.~G.}\
  \bibnamefont {Savvidis}},\ }\bibfield  {title} {\enquote {\bibinfo {title}
  {{All-dielectric GaN microcavity: Strong coupling and lasing at room
  temperature}},}\ }\href {\doibase 10.1063/1.4795019} {\bibfield  {journal}
  {\bibinfo  {journal} {Applied Physics Letters}\ }\textbf {\bibinfo {volume}
  {102}} (\bibinfo {year} {2013}),\ 10.1063/1.4795019}\BibitemShut {NoStop}%
\bibitem [{\citenamefont {Franke}\ \emph {et~al.}(2012)\citenamefont {Franke},
  \citenamefont {Sturm}, \citenamefont {Schmidt-Grund}, \citenamefont
  {Wagner},\ and\ \citenamefont {Grundmann}}]{Franke_2012}%
  \BibitemOpen
  \bibfield  {author} {\bibinfo {author} {\bibfnamefont {H.}~\bibnamefont
  {Franke}}, \bibinfo {author} {\bibfnamefont {C.}~\bibnamefont {Sturm}},
  \bibinfo {author} {\bibfnamefont {R.}~\bibnamefont {Schmidt-Grund}}, \bibinfo
  {author} {\bibfnamefont {G.}~\bibnamefont {Wagner}}, \ and\ \bibinfo {author}
  {\bibfnamefont {M.}~\bibnamefont {Grundmann}},\ }\bibfield  {title} {\enquote
  {\bibinfo {title} {{Ballistic propagation of exciton polariton condensates in
  a ZnO-based microcavity}},}\ }\href {\doibase 10.1088/1367-2630/14/1/013037}
  {\bibfield  {journal} {\bibinfo  {journal} {New Journal of Physics}\ }\textbf
  {\bibinfo {volume} {14}},\ \bibinfo {pages} {13037} (\bibinfo {year}
  {2012})}\BibitemShut {NoStop}%
\bibitem [{\citenamefont {Lidzey}\ \emph {et~al.}(1998)\citenamefont {Lidzey},
  \citenamefont {Bradley}, \citenamefont {Skolnick}, \citenamefont {Virgili},
  \citenamefont {Walker},\ and\ \citenamefont {Whittaker}}]{Lidzey1998}%
  \BibitemOpen
  \bibfield  {author} {\bibinfo {author} {\bibfnamefont {D.~G.}\ \bibnamefont
  {Lidzey}}, \bibinfo {author} {\bibfnamefont {D.~D.~C.}\ \bibnamefont
  {Bradley}}, \bibinfo {author} {\bibfnamefont {M.~S.}\ \bibnamefont
  {Skolnick}}, \bibinfo {author} {\bibfnamefont {T.}~\bibnamefont {Virgili}},
  \bibinfo {author} {\bibfnamefont {S.}~\bibnamefont {Walker}}, \ and\ \bibinfo
  {author} {\bibfnamefont {D.~M.}\ \bibnamefont {Whittaker}},\ }\bibfield
  {title} {\enquote {\bibinfo {title} {{Strong exciton–photon coupling in an
  organic semiconductor microcavity}},}\ }\href {\doibase 10.1038/25692}
  {\bibfield  {journal} {\bibinfo  {journal} {Nature}\ }\textbf {\bibinfo
  {volume} {395}},\ \bibinfo {pages} {53--55} (\bibinfo {year}
  {1998})}\BibitemShut {NoStop}%
\bibitem [{\citenamefont {Plumhof}\ \emph {et~al.}(2014)\citenamefont
  {Plumhof}, \citenamefont {St{\"{o}}ferle}, \citenamefont {Mai}, \citenamefont
  {Scherf},\ and\ \citenamefont {Mahrt}}]{Plumhof2014}%
  \BibitemOpen
  \bibfield  {author} {\bibinfo {author} {\bibfnamefont {J.~D.}\ \bibnamefont
  {Plumhof}}, \bibinfo {author} {\bibfnamefont {T.}~\bibnamefont
  {St{\"{o}}ferle}}, \bibinfo {author} {\bibfnamefont {L.}~\bibnamefont {Mai}},
  \bibinfo {author} {\bibfnamefont {U.}~\bibnamefont {Scherf}}, \ and\ \bibinfo
  {author} {\bibfnamefont {R.~F.}\ \bibnamefont {Mahrt}},\ }\bibfield  {title}
  {\enquote {\bibinfo {title} {{Room-temperature Bose–Einstein condensation
  of cavity exciton–polaritons in a polymer}},}\ }\href {\doibase
  10.1038/nmat3825} {\bibfield  {journal} {\bibinfo  {journal} {Nature
  Materials}\ }\textbf {\bibinfo {volume} {13}},\ \bibinfo {pages} {247--252}
  (\bibinfo {year} {2014})}\BibitemShut {NoStop}%
\bibitem [{\citenamefont {Liu}\ \emph {et~al.}(2014)\citenamefont {Liu},
  \citenamefont {Galfsky}, \citenamefont {Sun}, \citenamefont {Xia},
  \citenamefont {Lin}, \citenamefont {Lee}, \citenamefont {K{\'{e}}na-Cohen},\
  and\ \citenamefont {Menon}}]{Liu2014}%
  \BibitemOpen
  \bibfield  {author} {\bibinfo {author} {\bibfnamefont {X.}~\bibnamefont
  {Liu}}, \bibinfo {author} {\bibfnamefont {T.}~\bibnamefont {Galfsky}},
  \bibinfo {author} {\bibfnamefont {Z.}~\bibnamefont {Sun}}, \bibinfo {author}
  {\bibfnamefont {F.}~\bibnamefont {Xia}}, \bibinfo {author} {\bibfnamefont
  {E.-c.}\ \bibnamefont {Lin}}, \bibinfo {author} {\bibfnamefont {Y.-H.}\
  \bibnamefont {Lee}}, \bibinfo {author} {\bibfnamefont {S.}~\bibnamefont
  {K{\'{e}}na-Cohen}}, \ and\ \bibinfo {author} {\bibfnamefont {V.~M.}\
  \bibnamefont {Menon}},\ }\bibfield  {title} {\enquote {\bibinfo {title}
  {{Strong light–matter coupling in two-dimensional atomic crystals}},}\
  }\href {https://doi.org/10.1038/nphoton.2014.304
  http://10.0.4.14/nphoton.2014.304
  https://www.nature.com/articles/nphoton.2014.304{\#}supplementary-information}
  {\bibfield  {journal} {\bibinfo  {journal} {Nature Photonics}\ }\textbf
  {\bibinfo {volume} {9}},\ \bibinfo {pages} {30} (\bibinfo {year}
  {2014})}\BibitemShut {NoStop}%
\bibitem [{\citenamefont {Grosso}(2017)}]{Grosso2017}%
  \BibitemOpen
  \bibfield  {author} {\bibinfo {author} {\bibfnamefont {G.}~\bibnamefont
  {Grosso}},\ }\bibfield  {title} {\enquote {\bibinfo {title} {{2D materials:
  Valley polaritons}},}\ }\href {\doibase 10.1038/nphoton.2017.135} {\bibfield
  {journal} {\bibinfo  {journal} {Nature Photonics}\ }\textbf {\bibinfo
  {volume} {11}},\ \bibinfo {pages} {455--456} (\bibinfo {year}
  {2017})}\BibitemShut {NoStop}%
\bibitem [{\citenamefont {Fujita}\ \emph {et~al.}(1998)\citenamefont {Fujita},
  \citenamefont {Sato}, \citenamefont {Kuitani},\ and\ \citenamefont
  {Ishihara}}]{Fujita1998}%
  \BibitemOpen
  \bibfield  {author} {\bibinfo {author} {\bibfnamefont {T.}~\bibnamefont
  {Fujita}}, \bibinfo {author} {\bibfnamefont {Y.}~\bibnamefont {Sato}},
  \bibinfo {author} {\bibfnamefont {T.}~\bibnamefont {Kuitani}}, \ and\
  \bibinfo {author} {\bibfnamefont {T.}~\bibnamefont {Ishihara}},\ }\bibfield
  {title} {\enquote {\bibinfo {title} {{Tunable polariton absorption of
  distributed feedback microcavities at room temperature}},}\ }\href {\doibase
  10.1103/PhysRevB.57.12428} {\bibfield  {journal} {\bibinfo  {journal} {Phys.
  Rev. B}\ }\textbf {\bibinfo {volume} {57}},\ \bibinfo {pages} {12428--12434}
  (\bibinfo {year} {1998})}\BibitemShut {NoStop}%
\bibitem [{\citenamefont {Takada}, \citenamefont {Kamata},\ and\ \citenamefont
  {Bradley}(2003)}]{Takada2003}%
  \BibitemOpen
  \bibfield  {author} {\bibinfo {author} {\bibfnamefont {N.}~\bibnamefont
  {Takada}}, \bibinfo {author} {\bibfnamefont {T.}~\bibnamefont {Kamata}}, \
  and\ \bibinfo {author} {\bibfnamefont {D.~D.~C.}\ \bibnamefont {Bradley}},\
  }\bibfield  {title} {\enquote {\bibinfo {title} {{Polariton emission from
  polysilane-based organic microcavities}},}\ }\href {\doibase
  10.1063/1.1559950} {\bibfield  {journal} {\bibinfo  {journal} {Applied
  Physics Letters}\ }\textbf {\bibinfo {volume} {82}},\ \bibinfo {pages}
  {1812--1814} (\bibinfo {year} {2003})}\BibitemShut {NoStop}%
\bibitem [{\citenamefont {Lanty}\ \emph {et~al.}(2008)\citenamefont {Lanty},
  \citenamefont {Br{\'{e}}hier}, \citenamefont {Parashkov}, \citenamefont
  {Lauret},\ and\ \citenamefont {Deleporte}}]{lanty2008strong}%
  \BibitemOpen
  \bibfield  {author} {\bibinfo {author} {\bibfnamefont {G.}~\bibnamefont
  {Lanty}}, \bibinfo {author} {\bibfnamefont {A.}~\bibnamefont
  {Br{\'{e}}hier}}, \bibinfo {author} {\bibfnamefont {R.}~\bibnamefont
  {Parashkov}}, \bibinfo {author} {\bibfnamefont {J.-S.}\ \bibnamefont
  {Lauret}}, \ and\ \bibinfo {author} {\bibfnamefont {E.}~\bibnamefont
  {Deleporte}},\ }\bibfield  {title} {\enquote {\bibinfo {title} {{Strong
  exciton--photon coupling at room temperature in microcavities containing
  two-dimensional layered perovskite compounds}},}\ }\href
  {https://iopscience.iop.org/article/10.1088/1367-2630/10/6/065007} {\bibfield
   {journal} {\bibinfo  {journal} {New Journal of Physics}\ }\textbf {\bibinfo
  {volume} {10}},\ \bibinfo {pages} {65007} (\bibinfo {year}
  {2008})}\BibitemShut {NoStop}%
\bibitem [{\citenamefont {Nguyen}\ \emph {et~al.}(2014)\citenamefont {Nguyen},
  \citenamefont {Han}, \citenamefont {Abdel-Baki}, \citenamefont {Lafosse},
  \citenamefont {Amo}, \citenamefont {Lauret}, \citenamefont {Deleporte},
  \citenamefont {Bouchoule},\ and\ \citenamefont {Bloch}}]{Nguyen2014}%
  \BibitemOpen
  \bibfield  {author} {\bibinfo {author} {\bibfnamefont {H.~S.}\ \bibnamefont
  {Nguyen}}, \bibinfo {author} {\bibfnamefont {Z.}~\bibnamefont {Han}},
  \bibinfo {author} {\bibfnamefont {K.}~\bibnamefont {Abdel-Baki}}, \bibinfo
  {author} {\bibfnamefont {X.}~\bibnamefont {Lafosse}}, \bibinfo {author}
  {\bibfnamefont {A.}~\bibnamefont {Amo}}, \bibinfo {author} {\bibfnamefont
  {J.~S.}\ \bibnamefont {Lauret}}, \bibinfo {author} {\bibfnamefont
  {E.}~\bibnamefont {Deleporte}}, \bibinfo {author} {\bibfnamefont
  {S.}~\bibnamefont {Bouchoule}}, \ and\ \bibinfo {author} {\bibfnamefont
  {J.}~\bibnamefont {Bloch}},\ }\bibfield  {title} {\enquote {\bibinfo {title}
  {{Quantum confinement of zero-dimensional hybrid organic-inorganic polaritons
  at room temperature}},}\ }\href {\doibase 10.1063/1.4866606} {\bibfield
  {journal} {\bibinfo  {journal} {Applied Physics Letters}\ }\textbf {\bibinfo
  {volume} {104}},\ \bibinfo {pages} {1--5} (\bibinfo {year}
  {2014})}\BibitemShut {NoStop}%
\bibitem [{\citenamefont {Su}\ \emph {et~al.}(2017)\citenamefont {Su},
  \citenamefont {Diederichs}, \citenamefont {Wang}, \citenamefont {Liew},
  \citenamefont {Zhao}, \citenamefont {Liu}, \citenamefont {Xu}, \citenamefont
  {Chen},\ and\ \citenamefont {Xiong}}]{Su2017}%
  \BibitemOpen
  \bibfield  {author} {\bibinfo {author} {\bibfnamefont {R.}~\bibnamefont
  {Su}}, \bibinfo {author} {\bibfnamefont {C.}~\bibnamefont {Diederichs}},
  \bibinfo {author} {\bibfnamefont {J.}~\bibnamefont {Wang}}, \bibinfo {author}
  {\bibfnamefont {T.~C.~H.}\ \bibnamefont {Liew}}, \bibinfo {author}
  {\bibfnamefont {J.}~\bibnamefont {Zhao}}, \bibinfo {author} {\bibfnamefont
  {S.}~\bibnamefont {Liu}}, \bibinfo {author} {\bibfnamefont {W.}~\bibnamefont
  {Xu}}, \bibinfo {author} {\bibfnamefont {Z.}~\bibnamefont {Chen}}, \ and\
  \bibinfo {author} {\bibfnamefont {Q.}~\bibnamefont {Xiong}},\ }\bibfield
  {title} {\enquote {\bibinfo {title} {{Room-Temperature Polariton Lasing in
  All-Inorganic Perovskite Nanoplatelets}},}\ }\href {\doibase
  10.1021/acs.nanolett.7b01956} {\bibfield  {journal} {\bibinfo  {journal}
  {Nano Letters}\ }\textbf {\bibinfo {volume} {17}},\ \bibinfo {pages}
  {3982--3988} (\bibinfo {year} {2017})}\BibitemShut {NoStop}%
\bibitem [{\citenamefont {Fieramosca}\ \emph {et~al.}(2018)\citenamefont
  {Fieramosca}, \citenamefont {De~Marco}, \citenamefont {Passoni},
  \citenamefont {Polimeno}, \citenamefont {Rizzo}, \citenamefont {Rosa},
  \citenamefont {Cruciani}, \citenamefont {Dominici}, \citenamefont
  {De~Giorgi}, \citenamefont {Gigli}, \citenamefont {Andreani}, \citenamefont
  {Gerace}, \citenamefont {Ballarini},\ and\ \citenamefont
  {Sanvitto}}]{Fieramosca2018}%
  \BibitemOpen
  \bibfield  {author} {\bibinfo {author} {\bibfnamefont {A.}~\bibnamefont
  {Fieramosca}}, \bibinfo {author} {\bibfnamefont {L.}~\bibnamefont
  {De~Marco}}, \bibinfo {author} {\bibfnamefont {M.}~\bibnamefont {Passoni}},
  \bibinfo {author} {\bibfnamefont {L.}~\bibnamefont {Polimeno}}, \bibinfo
  {author} {\bibfnamefont {A.}~\bibnamefont {Rizzo}}, \bibinfo {author}
  {\bibfnamefont {B.~L.~T.}\ \bibnamefont {Rosa}}, \bibinfo {author}
  {\bibfnamefont {G.}~\bibnamefont {Cruciani}}, \bibinfo {author}
  {\bibfnamefont {L.}~\bibnamefont {Dominici}}, \bibinfo {author}
  {\bibfnamefont {M.}~\bibnamefont {De~Giorgi}}, \bibinfo {author}
  {\bibfnamefont {G.}~\bibnamefont {Gigli}}, \bibinfo {author} {\bibfnamefont
  {L.~C.}\ \bibnamefont {Andreani}}, \bibinfo {author} {\bibfnamefont
  {D.}~\bibnamefont {Gerace}}, \bibinfo {author} {\bibfnamefont
  {D.}~\bibnamefont {Ballarini}}, \ and\ \bibinfo {author} {\bibfnamefont
  {D.}~\bibnamefont {Sanvitto}},\ }\bibfield  {title} {\enquote {\bibinfo
  {title} {Tunable out-of-plane excitons in 2d single-crystal perovskites},}\
  }\href {\doibase 10.1021/acsphotonics.8b00984} {\bibfield  {journal}
  {\bibinfo  {journal} {ACS Photonics}\ }\textbf {\bibinfo {volume} {5}},\
  \bibinfo {pages} {4179--4185} (\bibinfo {year} {2018})},\ \Eprint
  {http://arxiv.org/abs/https://doi.org/10.1021/acsphotonics.8b00984}
  {https://doi.org/10.1021/acsphotonics.8b00984} \BibitemShut {NoStop}%
\bibitem [{\citenamefont {Su}\ \emph {et~al.}(2020)\citenamefont {Su},
  \citenamefont {Ghosh}, \citenamefont {Wang}, \citenamefont {Liu},
  \citenamefont {Diederichs}, \citenamefont {Liew},\ and\ \citenamefont
  {Xiong}}]{Su2020}%
  \BibitemOpen
  \bibfield  {author} {\bibinfo {author} {\bibfnamefont {R.}~\bibnamefont
  {Su}}, \bibinfo {author} {\bibfnamefont {S.}~\bibnamefont {Ghosh}}, \bibinfo
  {author} {\bibfnamefont {J.}~\bibnamefont {Wang}}, \bibinfo {author}
  {\bibfnamefont {S.}~\bibnamefont {Liu}}, \bibinfo {author} {\bibfnamefont
  {C.}~\bibnamefont {Diederichs}}, \bibinfo {author} {\bibfnamefont {T.~C.~H.}\
  \bibnamefont {Liew}}, \ and\ \bibinfo {author} {\bibfnamefont
  {Q.}~\bibnamefont {Xiong}},\ }\bibfield  {title} {\enquote {\bibinfo {title}
  {{Observation of exciton polariton condensation in a perovskite lattice at
  room temperature}},}\ }\href {\doibase 10.1038/s41567-019-0764-5} {\bibfield
  {journal} {\bibinfo  {journal} {Nature Physics}\ } (\bibinfo {year} {2020}),\
  10.1038/s41567-019-0764-5}\BibitemShut {NoStop}%
\bibitem [{\citenamefont {Dang}\ \emph {et~al.}(2020)\citenamefont {Dang},
  \citenamefont {Gerace}, \citenamefont {Drouard}, \citenamefont
  {Tripp{\'{e}}-Allard}, \citenamefont {L{\'{e}}d{\'{e}}e}, \citenamefont
  {Mazurczyk}, \citenamefont {Deleporte}, \citenamefont {Seassal},\ and\
  \citenamefont {Nguyen}}]{Dang2020}%
  \BibitemOpen
  \bibfield  {author} {\bibinfo {author} {\bibfnamefont {N.~H.~M.}\
  \bibnamefont {Dang}}, \bibinfo {author} {\bibfnamefont {D.}~\bibnamefont
  {Gerace}}, \bibinfo {author} {\bibfnamefont {E.}~\bibnamefont {Drouard}},
  \bibinfo {author} {\bibfnamefont {G.}~\bibnamefont {Tripp{\'{e}}-Allard}},
  \bibinfo {author} {\bibfnamefont {F.}~\bibnamefont {L{\'{e}}d{\'{e}}e}},
  \bibinfo {author} {\bibfnamefont {R.}~\bibnamefont {Mazurczyk}}, \bibinfo
  {author} {\bibfnamefont {E.}~\bibnamefont {Deleporte}}, \bibinfo {author}
  {\bibfnamefont {C.}~\bibnamefont {Seassal}}, \ and\ \bibinfo {author}
  {\bibfnamefont {H.~S.}\ \bibnamefont {Nguyen}},\ }\bibfield  {title}
  {\enquote {\bibinfo {title} {{Tailoring Dispersion of Room-Temperature
  Exciton-Polaritons with Perovskite-Based Subwavelength Metasurfaces}},}\
  }\href {\doibase 10.1021/acs.nanolett.0c00125} {\bibfield  {journal}
  {\bibinfo  {journal} {Nano Letters}\ }\textbf {\bibinfo {volume} {20}},\
  \bibinfo {pages} {2113--2119} (\bibinfo {year} {2020})}\BibitemShut {NoStop}%
\bibitem [{\citenamefont {Koshelev}\ \emph {et~al.}(2018)\citenamefont
  {Koshelev}, \citenamefont {Sychev}, \citenamefont {Sadrieva}, \citenamefont
  {Bogdanov},\ and\ \citenamefont {Iorsh}}]{Koshelev2018}%
  \BibitemOpen
  \bibfield  {author} {\bibinfo {author} {\bibfnamefont {K.~L.}\ \bibnamefont
  {Koshelev}}, \bibinfo {author} {\bibfnamefont {S.~K.}\ \bibnamefont
  {Sychev}}, \bibinfo {author} {\bibfnamefont {Z.~F.}\ \bibnamefont
  {Sadrieva}}, \bibinfo {author} {\bibfnamefont {A.~A.}\ \bibnamefont
  {Bogdanov}}, \ and\ \bibinfo {author} {\bibfnamefont {I.~V.}\ \bibnamefont
  {Iorsh}},\ }\bibfield  {title} {\enquote {\bibinfo {title} {Strong coupling
  between excitons in transition metal dichalcogenides and optical bound states
  in the continuum},}\ }\href {\doibase 10.1103/PhysRevB.98.161113} {\bibfield
  {journal} {\bibinfo  {journal} {Phys. Rev. B}\ }\textbf {\bibinfo {volume}
  {98}},\ \bibinfo {pages} {161113} (\bibinfo {year} {2018})}\BibitemShut
  {NoStop}%
\bibitem [{\citenamefont {Lu}\ \emph {et~al.}(2020)\citenamefont {Lu},
  \citenamefont {Le-Van}, \citenamefont {Ferrier}, \citenamefont {Drouard},
  \citenamefont {Seassal},\ and\ \citenamefont {Nguyen}}]{Lu:20}%
  \BibitemOpen
  \bibfield  {author} {\bibinfo {author} {\bibfnamefont {L.}~\bibnamefont
  {Lu}}, \bibinfo {author} {\bibfnamefont {Q.}~\bibnamefont {Le-Van}}, \bibinfo
  {author} {\bibfnamefont {L.}~\bibnamefont {Ferrier}}, \bibinfo {author}
  {\bibfnamefont {E.}~\bibnamefont {Drouard}}, \bibinfo {author} {\bibfnamefont
  {C.}~\bibnamefont {Seassal}}, \ and\ \bibinfo {author} {\bibfnamefont
  {H.~S.}\ \bibnamefont {Nguyen}},\ }\bibfield  {title} {\enquote {\bibinfo
  {title} {Engineering a light–matter strong coupling regime in
  perovskite-based plasmonic metasurface: quasi-bound state in the continuum
  and exceptional points},}\ }\href {\doibase 10.1364/PRJ.404743} {\bibfield
  {journal} {\bibinfo  {journal} {Photon. Res.}\ }\textbf {\bibinfo {volume}
  {8}},\ \bibinfo {pages} {A91--A100} (\bibinfo {year} {2020})}\BibitemShut
  {NoStop}%
\bibitem [{\citenamefont {Xie}\ \emph {et~al.}(2021)\citenamefont {Xie},
  \citenamefont {Liang}, \citenamefont {Jia}, \citenamefont {Li}, \citenamefont
  {Chen}, \citenamefont {Chang}, \citenamefont {Zhang},\ and\ \citenamefont
  {Wang}}]{Xie2021}%
  \BibitemOpen
  \bibfield  {author} {\bibinfo {author} {\bibfnamefont {P.}~\bibnamefont
  {Xie}}, \bibinfo {author} {\bibfnamefont {Z.}~\bibnamefont {Liang}}, \bibinfo
  {author} {\bibfnamefont {T.}~\bibnamefont {Jia}}, \bibinfo {author}
  {\bibfnamefont {D.}~\bibnamefont {Li}}, \bibinfo {author} {\bibfnamefont
  {Y.}~\bibnamefont {Chen}}, \bibinfo {author} {\bibfnamefont {P.}~\bibnamefont
  {Chang}}, \bibinfo {author} {\bibfnamefont {H.}~\bibnamefont {Zhang}}, \ and\
  \bibinfo {author} {\bibfnamefont {W.}~\bibnamefont {Wang}},\ }\bibfield
  {title} {\enquote {\bibinfo {title} {Strong coupling between excitons in a
  two-dimensional atomic crystal and quasibound states in the continuum in a
  two-dimensional all-dielectric asymmetric metasurface},}\ }\href {\doibase
  10.1103/PhysRevB.104.125446} {\bibfield  {journal} {\bibinfo  {journal}
  {Phys. Rev. B}\ }\textbf {\bibinfo {volume} {104}},\ \bibinfo {pages}
  {125446} (\bibinfo {year} {2021})}\BibitemShut {NoStop}%
\bibitem [{\citenamefont {Al-Ani}\ \emph {et~al.}()\citenamefont {Al-Ani},
  \citenamefont {As'Ham}, \citenamefont {Huang}, \citenamefont
  {Miroshnichenko}, \citenamefont {Lei},\ and\ \citenamefont
  {Hattori}}]{Ani2021}%
  \BibitemOpen
  \bibfield  {author} {\bibinfo {author} {\bibfnamefont {I.~A.~M.}\
  \bibnamefont {Al-Ani}}, \bibinfo {author} {\bibfnamefont {K.}~\bibnamefont
  {As'Ham}}, \bibinfo {author} {\bibfnamefont {L.}~\bibnamefont {Huang}},
  \bibinfo {author} {\bibfnamefont {A.~E.}\ \bibnamefont {Miroshnichenko}},
  \bibinfo {author} {\bibfnamefont {W.}~\bibnamefont {Lei}}, \ and\ \bibinfo
  {author} {\bibfnamefont {H.~T.}\ \bibnamefont {Hattori}},\ }\bibfield
  {title} {\enquote {\bibinfo {title} {Strong coupling of exciton and high-q
  mode in all-perovskite metasurfaces},}\ }\href {\doibase
  https://doi.org/10.1002/adom.202101120} {\bibfield  {journal} {\bibinfo
  {journal} {Advanced Optical Materials}\ }\textbf {\bibinfo {volume} {n/a}},\
  \bibinfo {pages} {2101120}},\ \Eprint
  {http://arxiv.org/abs/https://onlinelibrary.wiley.com/doi/pdf/10.1002/adom.202101120}
  {https://onlinelibrary.wiley.com/doi/pdf/10.1002/adom.202101120} \BibitemShut
  {NoStop}%
\bibitem [{\citenamefont {Kravtsov}\ \emph {et~al.}(2020)\citenamefont
  {Kravtsov}, \citenamefont {Khestanova}, \citenamefont {Benimetskiy},
  \citenamefont {Ivanova}, \citenamefont {Samusev}, \citenamefont {Sinev},
  \citenamefont {Pidgayko}, \citenamefont {Mozharov}, \citenamefont {Mukhin},
  \citenamefont {Lozhkin}, \citenamefont {Kapitonov}, \citenamefont {Brichkin},
  \citenamefont {Kulakovskii}, \citenamefont {Shelykh}, \citenamefont
  {Tartakovskii}, \citenamefont {Walker}, \citenamefont {Skolnick},
  \citenamefont {Krizhanovskii},\ and\ \citenamefont {Iorsh}}]{Kravtsov2020}%
  \BibitemOpen
  \bibfield  {author} {\bibinfo {author} {\bibfnamefont {V.}~\bibnamefont
  {Kravtsov}}, \bibinfo {author} {\bibfnamefont {E.}~\bibnamefont
  {Khestanova}}, \bibinfo {author} {\bibfnamefont {F.~A.}\ \bibnamefont
  {Benimetskiy}}, \bibinfo {author} {\bibfnamefont {T.}~\bibnamefont
  {Ivanova}}, \bibinfo {author} {\bibfnamefont {A.~K.}\ \bibnamefont
  {Samusev}}, \bibinfo {author} {\bibfnamefont {I.~S.}\ \bibnamefont {Sinev}},
  \bibinfo {author} {\bibfnamefont {D.}~\bibnamefont {Pidgayko}}, \bibinfo
  {author} {\bibfnamefont {A.~M.}\ \bibnamefont {Mozharov}}, \bibinfo {author}
  {\bibfnamefont {I.~S.}\ \bibnamefont {Mukhin}}, \bibinfo {author}
  {\bibfnamefont {M.~S.}\ \bibnamefont {Lozhkin}}, \bibinfo {author}
  {\bibfnamefont {Y.~V.}\ \bibnamefont {Kapitonov}}, \bibinfo {author}
  {\bibfnamefont {A.~S.}\ \bibnamefont {Brichkin}}, \bibinfo {author}
  {\bibfnamefont {V.~D.}\ \bibnamefont {Kulakovskii}}, \bibinfo {author}
  {\bibfnamefont {I.~A.}\ \bibnamefont {Shelykh}}, \bibinfo {author}
  {\bibfnamefont {A.~I.}\ \bibnamefont {Tartakovskii}}, \bibinfo {author}
  {\bibfnamefont {P.~M.}\ \bibnamefont {Walker}}, \bibinfo {author}
  {\bibfnamefont {M.~S.}\ \bibnamefont {Skolnick}}, \bibinfo {author}
  {\bibfnamefont {D.~N.}\ \bibnamefont {Krizhanovskii}}, \ and\ \bibinfo
  {author} {\bibfnamefont {I.~V.}\ \bibnamefont {Iorsh}},\ }\bibfield  {title}
  {\enquote {\bibinfo {title} {{Nonlinear polaritons in a monolayer
  semiconductor coupled to optical bound states in the continuum}},}\ }\href
  {\doibase 10.1038/s41377-020-0286-z} {\bibfield  {journal} {\bibinfo
  {journal} {Light: Science \& Applications}\ }\textbf {\bibinfo {volume}
  {9}},\ \bibinfo {pages} {56} (\bibinfo {year} {2020})}\BibitemShut {NoStop}%
\bibitem [{\citenamefont {Ardizzone}\ \emph {et~al.}(2021)\citenamefont
  {Ardizzone}, \citenamefont {Riminucci}, \citenamefont {Zanotti},
  \citenamefont {Gianfrate}, \citenamefont {Suarez-Forero}, \citenamefont
  {Todisco}, \citenamefont {Giorgi}, \citenamefont {Trypogeorgos},
  \citenamefont {Gigli}, \citenamefont {Nguyen}, \citenamefont {Baldwin},
  \citenamefont {Pfeiffer}, \citenamefont {Ballarini}, \citenamefont {Gerace},\
  and\ \citenamefont {Sanvitto}}]{ardizzone2021polariton}%
  \BibitemOpen
  \bibfield  {author} {\bibinfo {author} {\bibfnamefont {V.}~\bibnamefont
  {Ardizzone}}, \bibinfo {author} {\bibfnamefont {F.}~\bibnamefont
  {Riminucci}}, \bibinfo {author} {\bibfnamefont {S.}~\bibnamefont {Zanotti}},
  \bibinfo {author} {\bibfnamefont {A.}~\bibnamefont {Gianfrate}}, \bibinfo
  {author} {\bibfnamefont {D.~G.}\ \bibnamefont {Suarez-Forero}}, \bibinfo
  {author} {\bibfnamefont {F.}~\bibnamefont {Todisco}}, \bibinfo {author}
  {\bibfnamefont {M.~D.}\ \bibnamefont {Giorgi}}, \bibinfo {author}
  {\bibfnamefont {D.}~\bibnamefont {Trypogeorgos}}, \bibinfo {author}
  {\bibfnamefont {G.}~\bibnamefont {Gigli}}, \bibinfo {author} {\bibfnamefont
  {H.~S.}\ \bibnamefont {Nguyen}}, \bibinfo {author} {\bibfnamefont
  {K.}~\bibnamefont {Baldwin}}, \bibinfo {author} {\bibfnamefont
  {L.}~\bibnamefont {Pfeiffer}}, \bibinfo {author} {\bibfnamefont
  {D.}~\bibnamefont {Ballarini}}, \bibinfo {author} {\bibfnamefont
  {D.}~\bibnamefont {Gerace}}, \ and\ \bibinfo {author} {\bibfnamefont
  {D.}~\bibnamefont {Sanvitto}},\ }\href@noop {} {\enquote {\bibinfo {title}
  {Polariton bose-einstein condensate from a bound state in the continuum},}\ }
  (\bibinfo {year} {2021}),\ \Eprint {http://arxiv.org/abs/2105.09362}
  {arXiv:2105.09362 [physics.optics]} \BibitemShut {NoStop}%
\bibitem [{\citenamefont {Su}\ \emph {et~al.}(2021)\citenamefont {Su},
  \citenamefont {Fieramosca}, \citenamefont {Zhang}, \citenamefont {Nguyen},
  \citenamefont {Deleporte}, \citenamefont {Chen}, \citenamefont {Sanvitto},
  \citenamefont {Liew},\ and\ \citenamefont {Xiong}}]{Su2021}%
  \BibitemOpen
  \bibfield  {author} {\bibinfo {author} {\bibfnamefont {R.}~\bibnamefont
  {Su}}, \bibinfo {author} {\bibfnamefont {A.}~\bibnamefont {Fieramosca}},
  \bibinfo {author} {\bibfnamefont {Q.}~\bibnamefont {Zhang}}, \bibinfo
  {author} {\bibfnamefont {H.~S.}\ \bibnamefont {Nguyen}}, \bibinfo {author}
  {\bibfnamefont {E.}~\bibnamefont {Deleporte}}, \bibinfo {author}
  {\bibfnamefont {Z.}~\bibnamefont {Chen}}, \bibinfo {author} {\bibfnamefont
  {D.}~\bibnamefont {Sanvitto}}, \bibinfo {author} {\bibfnamefont {T.~C.~H.}\
  \bibnamefont {Liew}}, \ and\ \bibinfo {author} {\bibfnamefont
  {Q.}~\bibnamefont {Xiong}},\ }\bibfield  {title} {\enquote {\bibinfo {title}
  {{Perovskite semiconductors for room-temperature exciton-polaritonics}},}\
  }\href {\doibase 10.1038/s41563-021-01035-x} {\bibfield  {journal} {\bibinfo
  {journal} {Nature Materials}\ } (\bibinfo {year} {2021}),\
  10.1038/s41563-021-01035-x}\BibitemShut {NoStop}%
\bibitem [{\citenamefont {Gauthron}\ \emph {et~al.}(2010)\citenamefont
  {Gauthron}, \citenamefont {Lauret}, \citenamefont {Doyennette}, \citenamefont
  {Lanty}, \citenamefont {Choueiry}, \citenamefont {Zhang}, \citenamefont
  {Brehier}, \citenamefont {Largeau}, \citenamefont {Mauguin}, \citenamefont
  {Bloch},\ and\ \citenamefont {Deleporte}}]{Gauthron2010}%
  \BibitemOpen
  \bibfield  {author} {\bibinfo {author} {\bibfnamefont {K.}~\bibnamefont
  {Gauthron}}, \bibinfo {author} {\bibfnamefont {J.-S.}\ \bibnamefont
  {Lauret}}, \bibinfo {author} {\bibfnamefont {L.}~\bibnamefont {Doyennette}},
  \bibinfo {author} {\bibfnamefont {G.}~\bibnamefont {Lanty}}, \bibinfo
  {author} {\bibfnamefont {A.~A.}\ \bibnamefont {Choueiry}}, \bibinfo {author}
  {\bibfnamefont {S.~J.}\ \bibnamefont {Zhang}}, \bibinfo {author}
  {\bibfnamefont {A.}~\bibnamefont {Brehier}}, \bibinfo {author} {\bibfnamefont
  {L.}~\bibnamefont {Largeau}}, \bibinfo {author} {\bibfnamefont
  {O.}~\bibnamefont {Mauguin}}, \bibinfo {author} {\bibfnamefont
  {J.}~\bibnamefont {Bloch}}, \ and\ \bibinfo {author} {\bibfnamefont
  {E.}~\bibnamefont {Deleporte}},\ }\bibfield  {title} {\enquote {\bibinfo
  {title} {{Optical spectroscopy of two-dimensional layered
  (C6H5C2H4-NH3)2-PbI4 perovskite}},}\ }\href {\doibase 10.1364/OE.18.005912}
  {\bibfield  {journal} {\bibinfo  {journal} {Opt. Express}\ }\textbf {\bibinfo
  {volume} {18}},\ \bibinfo {pages} {5912--5919} (\bibinfo {year}
  {2010})}\BibitemShut {NoStop}%
\bibitem [{\citenamefont {Andreani}\ and\ \citenamefont
  {Gerace}(2006)}]{Andreani-Gerace2006}%
  \BibitemOpen
  \bibfield  {author} {\bibinfo {author} {\bibfnamefont {L.~C.}\ \bibnamefont
  {Andreani}}\ and\ \bibinfo {author} {\bibfnamefont {D.}~\bibnamefont
  {Gerace}},\ }\bibfield  {title} {\enquote {\bibinfo {title} {Photonic-crystal
  slabs with a triangular lattice of triangular holes investigated using a
  guided-mode expansion method},}\ }\href {\doibase 10.1103/PhysRevB.73.235114}
  {\bibfield  {journal} {\bibinfo  {journal} {Phys. Rev. B}\ }\textbf {\bibinfo
  {volume} {73}},\ \bibinfo {pages} {235114} (\bibinfo {year}
  {2006})}\BibitemShut {NoStop}%
\bibitem [{\citenamefont {Minkov}\ \emph {et~al.}(2020)\citenamefont {Minkov},
  \citenamefont {Williamson}, \citenamefont {Andreani}, \citenamefont {Gerace},
  \citenamefont {Lou}, \citenamefont {Song}, \citenamefont {Hughes},\ and\
  \citenamefont {Fan}}]{Minkov2020}%
  \BibitemOpen
  \bibfield  {author} {\bibinfo {author} {\bibfnamefont {M.}~\bibnamefont
  {Minkov}}, \bibinfo {author} {\bibfnamefont {I.~A.~D.}\ \bibnamefont
  {Williamson}}, \bibinfo {author} {\bibfnamefont {L.~C.}\ \bibnamefont
  {Andreani}}, \bibinfo {author} {\bibfnamefont {D.}~\bibnamefont {Gerace}},
  \bibinfo {author} {\bibfnamefont {B.}~\bibnamefont {Lou}}, \bibinfo {author}
  {\bibfnamefont {A.~Y.}\ \bibnamefont {Song}}, \bibinfo {author}
  {\bibfnamefont {T.~W.}\ \bibnamefont {Hughes}}, \ and\ \bibinfo {author}
  {\bibfnamefont {S.}~\bibnamefont {Fan}},\ }\bibfield  {title} {\enquote
  {\bibinfo {title} {Inverse design of photonic crystals through automatic
  differentiation},}\ }\href {\doibase 10.1021/acsphotonics.0c00327} {\bibfield
   {journal} {\bibinfo  {journal} {ACS Photonics}\ }\textbf {\bibinfo {volume}
  {7}},\ \bibinfo {pages} {1729--1741} (\bibinfo {year} {2020})}\BibitemShut
  {NoStop}%
\bibitem [{\citenamefont {Gerace}\ and\ \citenamefont
  {Andreani}(2007)}]{Gerace-Andreani2007}%
  \BibitemOpen
  \bibfield  {author} {\bibinfo {author} {\bibfnamefont {D.}~\bibnamefont
  {Gerace}}\ and\ \bibinfo {author} {\bibfnamefont {L.~C.}\ \bibnamefont
  {Andreani}},\ }\bibfield  {title} {\enquote {\bibinfo {title} {Quantum theory
  of exciton-photon coupling in photonic crystal slabs with embedded quantum
  wells},}\ }\href {\doibase 10.1103/PhysRevB.75.235325} {\bibfield  {journal}
  {\bibinfo  {journal} {Phys. Rev. B}\ }\textbf {\bibinfo {volume} {75}},\
  \bibinfo {pages} {235325} (\bibinfo {year} {2007})}\BibitemShut {NoStop}%
\bibitem [{\citenamefont {Auff{\`e}ves}\ \emph {et~al.}(2014)\citenamefont
  {Auff{\`e}ves}, \citenamefont {Gerace}, \citenamefont {Richard},
  \citenamefont {Portolan}, \citenamefont {Fran{\c c}a~Santos}, \citenamefont
  {Kwek},\ and\ \citenamefont {Miniatura}}]{StrongCoupling_book}%
  \BibitemOpen
  \bibfield  {author} {\bibinfo {author} {\bibfnamefont {A.}~\bibnamefont
  {Auff{\`e}ves}}, \bibinfo {author} {\bibfnamefont {D.}~\bibnamefont
  {Gerace}}, \bibinfo {author} {\bibfnamefont {M.}~\bibnamefont {Richard}},
  \bibinfo {author} {\bibfnamefont {S.}~\bibnamefont {Portolan}}, \bibinfo
  {author} {\bibfnamefont {M.}~\bibnamefont {Fran{\c c}a~Santos}}, \bibinfo
  {author} {\bibfnamefont {L.~C.}\ \bibnamefont {Kwek}}, \ and\ \bibinfo
  {author} {\bibfnamefont {C.}~\bibnamefont {Miniatura}},\ }\href {\doibase
  10.1142/8758} {\emph {\bibinfo {title} {Strong Light-Matter Coupling}}}\
  (\bibinfo  {publisher} {WORLD SCIENTIFIC},\ \bibinfo {year} {2014})\ \Eprint
  {http://arxiv.org/abs/https://www.worldscientific.com/doi/pdf/10.1142/8758}
  {https://www.worldscientific.com/doi/pdf/10.1142/8758} \BibitemShut {NoStop}%
\bibitem [{\citenamefont {Galli}\ \emph {et~al.}(2005)\citenamefont {Galli},
  \citenamefont {Bajoni}, \citenamefont {Belotti}, \citenamefont {Paleari},
  \citenamefont {Patrini}, \citenamefont {Guizzetti}, \citenamefont {Gerace},
  \citenamefont {Agio}, \citenamefont {Andreani}, \citenamefont {Peyrade},\
  and\ \citenamefont {Chen}}]{Galli2005}%
  \BibitemOpen
  \bibfield  {author} {\bibinfo {author} {\bibfnamefont {M.}~\bibnamefont
  {Galli}}, \bibinfo {author} {\bibfnamefont {D.}~\bibnamefont {Bajoni}},
  \bibinfo {author} {\bibfnamefont {M.}~\bibnamefont {Belotti}}, \bibinfo
  {author} {\bibfnamefont {F.}~\bibnamefont {Paleari}}, \bibinfo {author}
  {\bibfnamefont {M.}~\bibnamefont {Patrini}}, \bibinfo {author} {\bibfnamefont
  {G.}~\bibnamefont {Guizzetti}}, \bibinfo {author} {\bibfnamefont
  {D.}~\bibnamefont {Gerace}}, \bibinfo {author} {\bibfnamefont
  {M.}~\bibnamefont {Agio}}, \bibinfo {author} {\bibfnamefont {L.}~\bibnamefont
  {Andreani}}, \bibinfo {author} {\bibfnamefont {D.}~\bibnamefont {Peyrade}}, \
  and\ \bibinfo {author} {\bibfnamefont {Y.}~\bibnamefont {Chen}},\ }\bibfield
  {title} {\enquote {\bibinfo {title} {Measurement of photonic mode dispersion
  and linewidths in silicon-on-insulator photonic crystal slabs},}\ }\href
  {\doibase 10.1109/JSAC.2005.851164} {\bibfield  {journal} {\bibinfo
  {journal} {IEEE Journal on Selected Areas in Communications}\ }\textbf
  {\bibinfo {volume} {23}},\ \bibinfo {pages} {1402--1410} (\bibinfo {year}
  {2005})}\BibitemShut {NoStop}%
\bibitem [{\citenamefont {McCutcheon}\ \emph {et~al.}(2005)\citenamefont
  {McCutcheon}, \citenamefont {Rieger}, \citenamefont {Cheung}, \citenamefont
  {Young}, \citenamefont {Dalacu}, \citenamefont {Fr{\'e}d{\'e}rick},
  \citenamefont {Poole}, \citenamefont {Aers},\ and\ \citenamefont
  {Williams}}]{McCutcheon2005}%
  \BibitemOpen
  \bibfield  {author} {\bibinfo {author} {\bibfnamefont {M.~W.}\ \bibnamefont
  {McCutcheon}}, \bibinfo {author} {\bibfnamefont {G.~W.}\ \bibnamefont
  {Rieger}}, \bibinfo {author} {\bibfnamefont {I.~W.}\ \bibnamefont {Cheung}},
  \bibinfo {author} {\bibfnamefont {J.~F.}\ \bibnamefont {Young}}, \bibinfo
  {author} {\bibfnamefont {D.}~\bibnamefont {Dalacu}}, \bibinfo {author}
  {\bibfnamefont {S.}~\bibnamefont {Fr{\'e}d{\'e}rick}}, \bibinfo {author}
  {\bibfnamefont {P.~J.}\ \bibnamefont {Poole}}, \bibinfo {author}
  {\bibfnamefont {G.~C.}\ \bibnamefont {Aers}}, \ and\ \bibinfo {author}
  {\bibfnamefont {R.~L.}\ \bibnamefont {Williams}},\ }\bibfield  {title}
  {\enquote {\bibinfo {title} {Resonant scattering and second-harmonic
  spectroscopy of planar photonic crystal microcavities},}\ }\href {\doibase
  10.1063/1.2137898} {\bibfield  {journal} {\bibinfo  {journal} {Applied
  Physics Letters}\ }\textbf {\bibinfo {volume} {87}},\ \bibinfo {pages}
  {221110} (\bibinfo {year} {2005})},\ \Eprint
  {http://arxiv.org/abs/https://doi.org/10.1063/1.2137898}
  {https://doi.org/10.1063/1.2137898} \BibitemShut {NoStop}%
\bibitem [{\citenamefont {Liu}\ and\ \citenamefont {Fan}(2012)}]{Liu2012}%
  \BibitemOpen
  \bibfield  {author} {\bibinfo {author} {\bibfnamefont {V.}~\bibnamefont
  {Liu}}\ and\ \bibinfo {author} {\bibfnamefont {S.}~\bibnamefont {Fan}},\
  }\bibfield  {title} {\enquote {\bibinfo {title} {{S4 : A free electromagnetic
  solver for layered periodic structures}},}\ }\href {\doibase
  10.1016/J.CPC.2012.04.026} {\bibfield  {journal} {\bibinfo  {journal}
  {Computer Physics Communications}\ }\textbf {\bibinfo {volume} {183}},\
  \bibinfo {pages} {2233--2244} (\bibinfo {year} {2012})}\BibitemShut {NoStop}%
\bibitem [{\citenamefont {Hsu}\ \emph {et~al.}(2013)\citenamefont {Hsu},
  \citenamefont {Zhen}, \citenamefont {Lee}, \citenamefont {Chua},
  \citenamefont {Johnson}, \citenamefont {Joannopoulos},\ and\ \citenamefont
  {Solja{\v{c}}i{\'{c}}}}]{Hsu2013}%
  \BibitemOpen
  \bibfield  {author} {\bibinfo {author} {\bibfnamefont {C.~W.}\ \bibnamefont
  {Hsu}}, \bibinfo {author} {\bibfnamefont {B.}~\bibnamefont {Zhen}}, \bibinfo
  {author} {\bibfnamefont {J.}~\bibnamefont {Lee}}, \bibinfo {author}
  {\bibfnamefont {S.-l.}\ \bibnamefont {Chua}}, \bibinfo {author}
  {\bibfnamefont {S.~G.}\ \bibnamefont {Johnson}}, \bibinfo {author}
  {\bibfnamefont {J.~D.}\ \bibnamefont {Joannopoulos}}, \ and\ \bibinfo
  {author} {\bibfnamefont {M.}~\bibnamefont {Solja{\v{c}}i{\'{c}}}},\
  }\bibfield  {title} {\enquote {\bibinfo {title} {{Observation of trapped
  light within the radiation continuum}},}\ }\href {\doibase
  10.1038/nature12289} {\bibfield  {journal} {\bibinfo  {journal} {Nature}\
  }\textbf {\bibinfo {volume} {499}},\ \bibinfo {pages} {188--191} (\bibinfo
  {year} {2013})}\BibitemShut {NoStop}%
\end{thebibliography}%
	
\newpage 
	
\onecolumngrid
\appendix

\newpage
\begin{center}
	\textbf{\large --- Supplementary Material ---\\} 
\end{center}
\setcounter{equation}{0}
\setcounter{figure}{0}
\setcounter{table}{0}
\setcounter{page}{1}
\makeatletter
\renewcommand{\thesection}{S\arabic{section}}
\renewcommand{\theequation}{S\arabic{section}.\arabic{equation}}
\renewcommand{\thefigure}{S\arabic{figure}}
\renewcommand{\bibnumfmt}[1]{[S#1]}
		
	\begin{figure*}[htb]
		\begin{center}
			\includegraphics[width=0.7\textwidth]{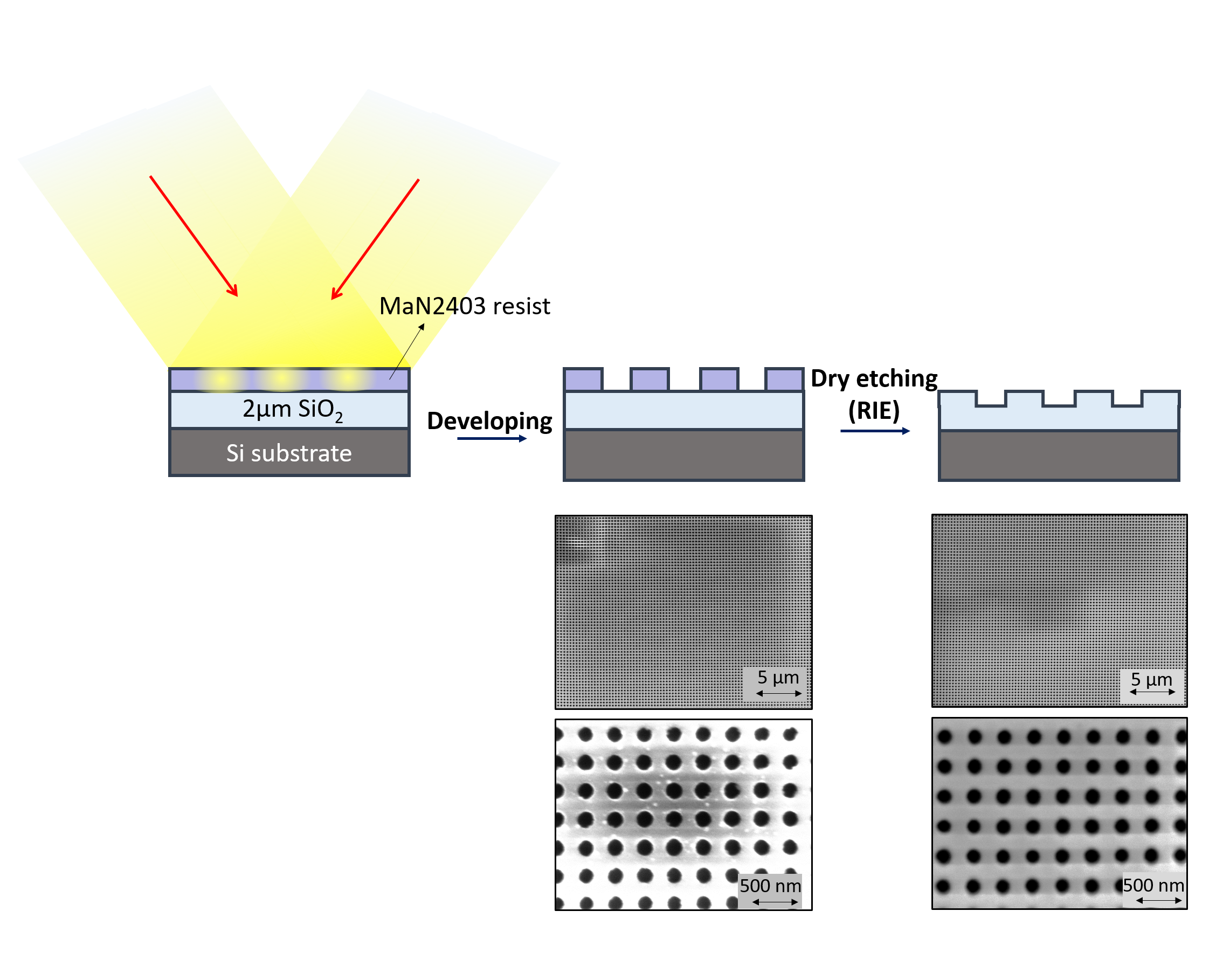}
		\end{center}
		\caption{\label{figS1}LIL fabrication scheme of SiO$_2$ backbones and SEM images of samples before and after RIE process.}
	\end{figure*}
	
	\begin{figure*}[htb]
		\begin{center}
			\includegraphics[width=0.6\textwidth]{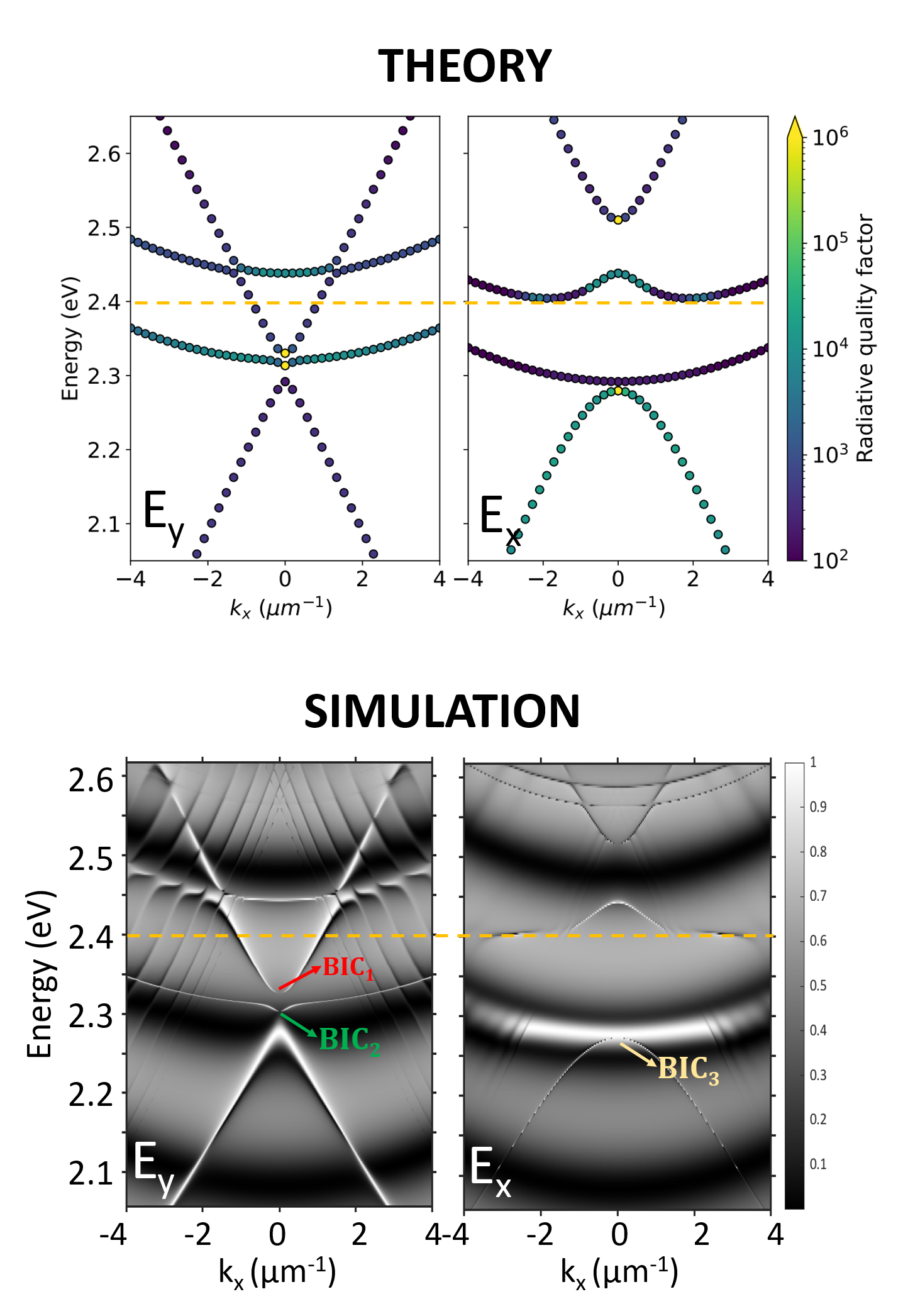}
		\end{center}
		\caption{\label{figS2}{Comparison between theoretical band structure of the passive PEPI metasurface, calculated by using the GME method (top panels) for both odd and even parity with respect to the $xz$ plane, and the angle-resolved reflectivity spectra simulated by RCWA method (lower panels). Here, the considered structure is the passive one ($A_X=0$). Interestingly, other than BIC$_1$ and BIC$_2$ in $E_y$ polarization (i.e. odd parity about $xz$ plane), other photonic BICs (here denoted BIC$_3$ and BIC$_4$) exist but they can be excited in $E_x$ polarization (i.e., even parity about $xz$ plane). However, they are far from the excitonic resonances (BIC$_3$ is at lower frequency and BIC$_4$ is at higher frequency), and their strong coupling with the excitonic resonance result in very photonic pol-BICs, and they are not populated at all in photoluminescence measurements.   }}
	\end{figure*}
		
	\begin{figure*}[htb]
		\begin{center}
			\includegraphics[width=0.8\textwidth]{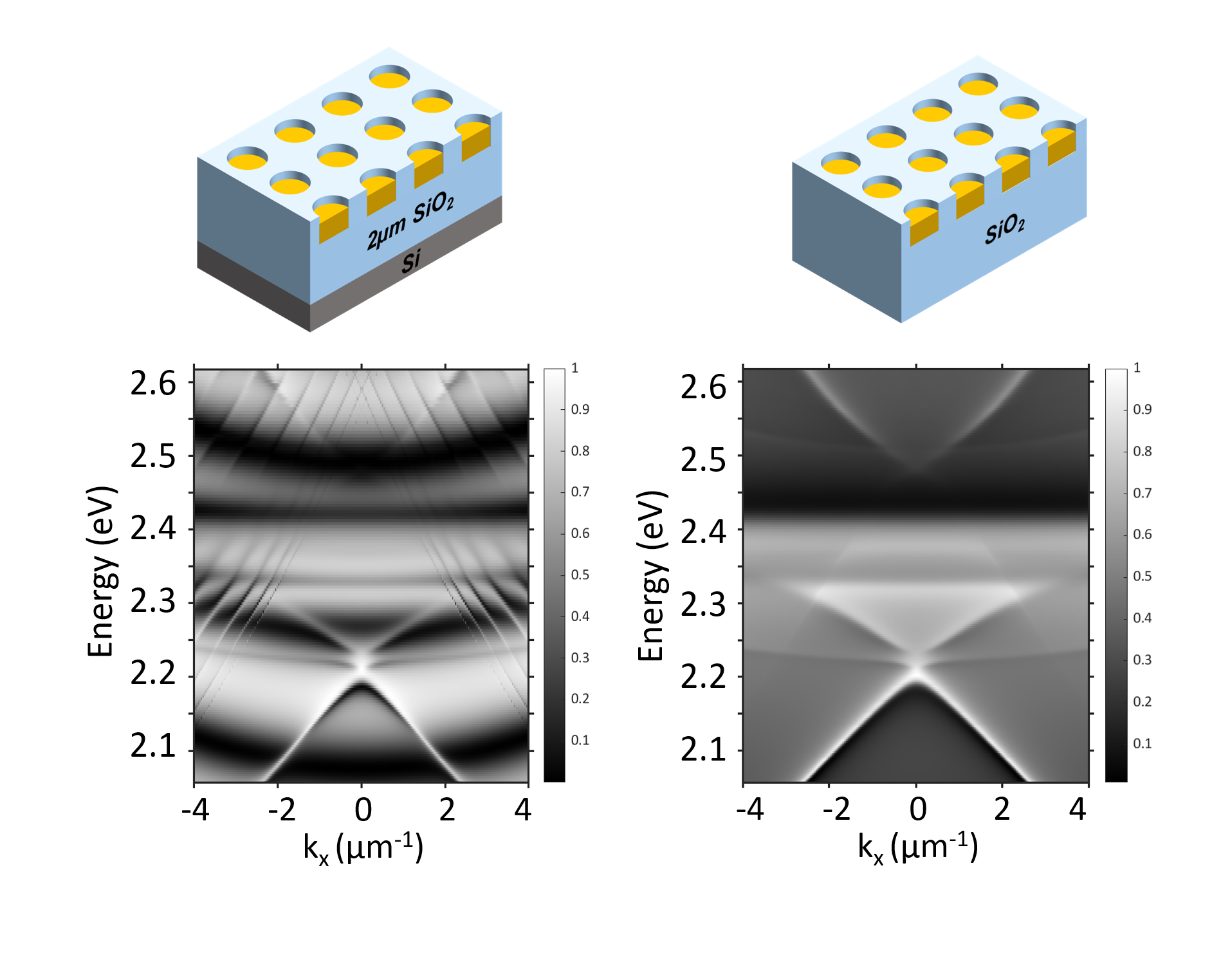}
		\end{center}
		\caption{\label{figS3}Comparison between angle-resolved reflectivity spectra simulated by RCWA method of PEPI infiltrated inside 2$\mu$m SiO$_2$ on Si substrate (left panel) and inside pure SiO$_2$ substrate (right panel). These results allow to confirm which resonances come from the metasurface and not from the confinement within the SiO$_2$ layer.}
	\end{figure*}
	
	\begin{figure*}[htb]
		\begin{center}
			\includegraphics[width=0.7\textwidth]{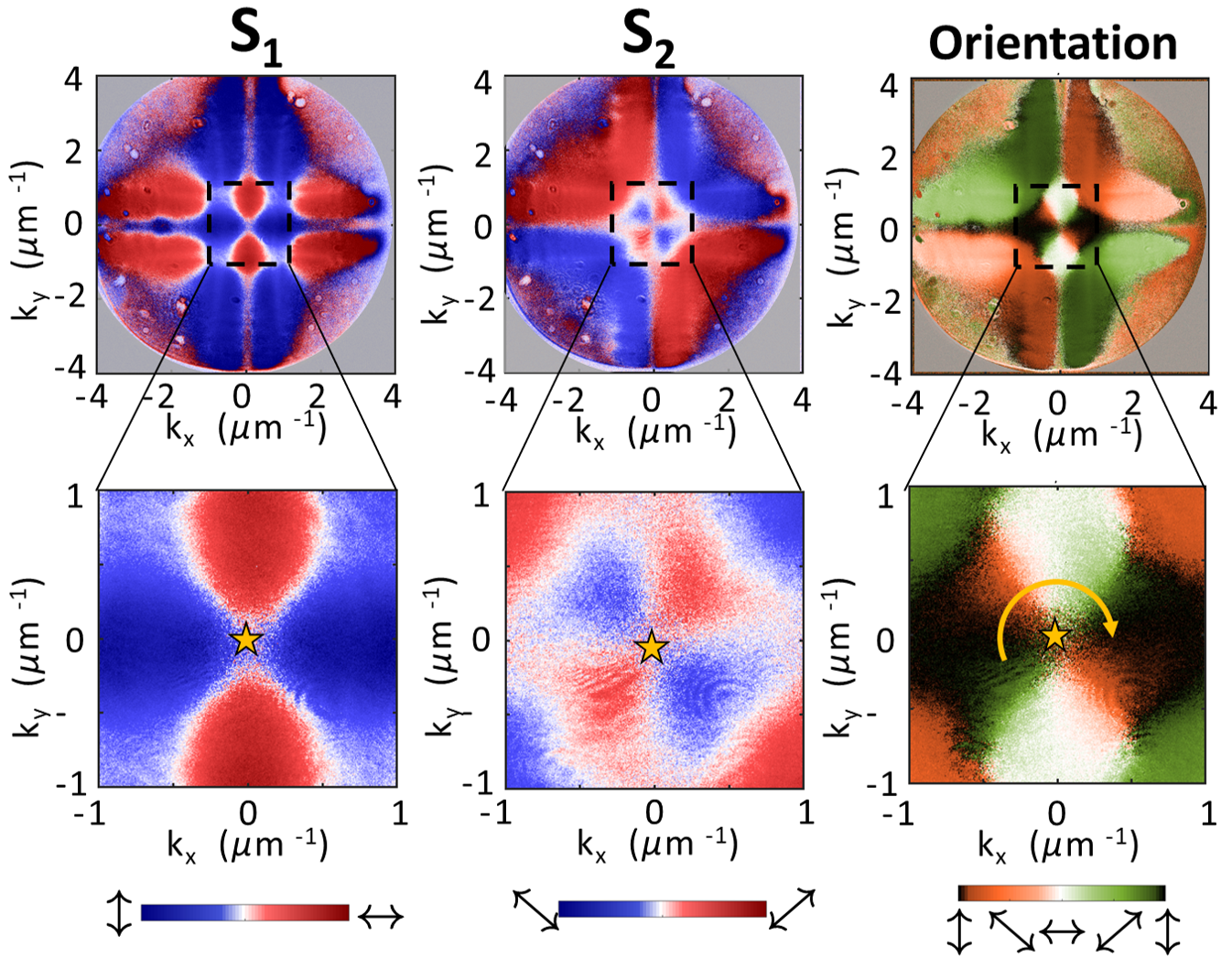}
		\end{center}
		\caption{\label{figS4}Experimental results of the Stokes parameters, $S_1$ and $S_2$, and the corresponding orientation angle, $\phi$. Top panels: Unzoomed data. Lower panels: Zoomed data (as shown in the main text). }
	\end{figure*}
	
	\begin{figure*}[htb]
		\begin{center}
			\includegraphics[width=0.8\textwidth]{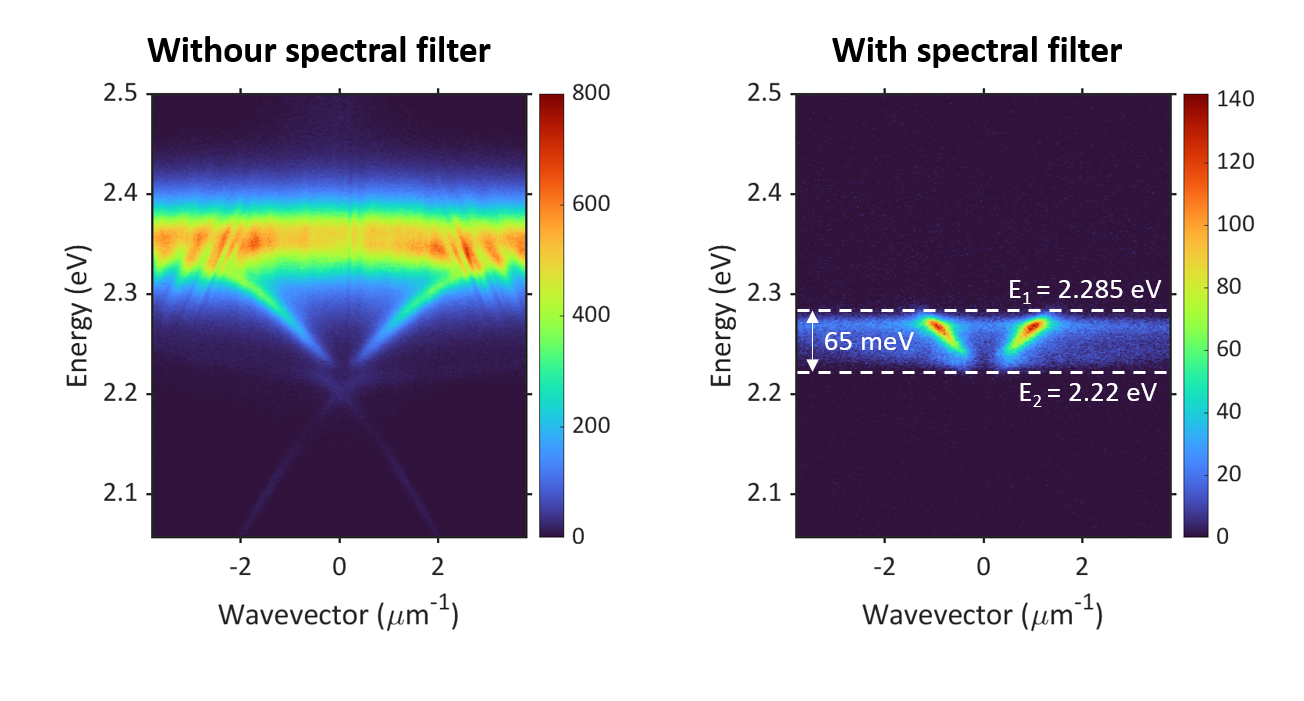}
		\end{center}
		\caption{\label{figS5}Experimental angle-resolved photoluminescence dispersion with and without spectral filter. Here the spectral filter is the one used for mapping the Stoke parameters.}
	\end{figure*}
	
	\begin{figure*}[htb]
		\begin{center}
			\includegraphics[width=0.8\textwidth]{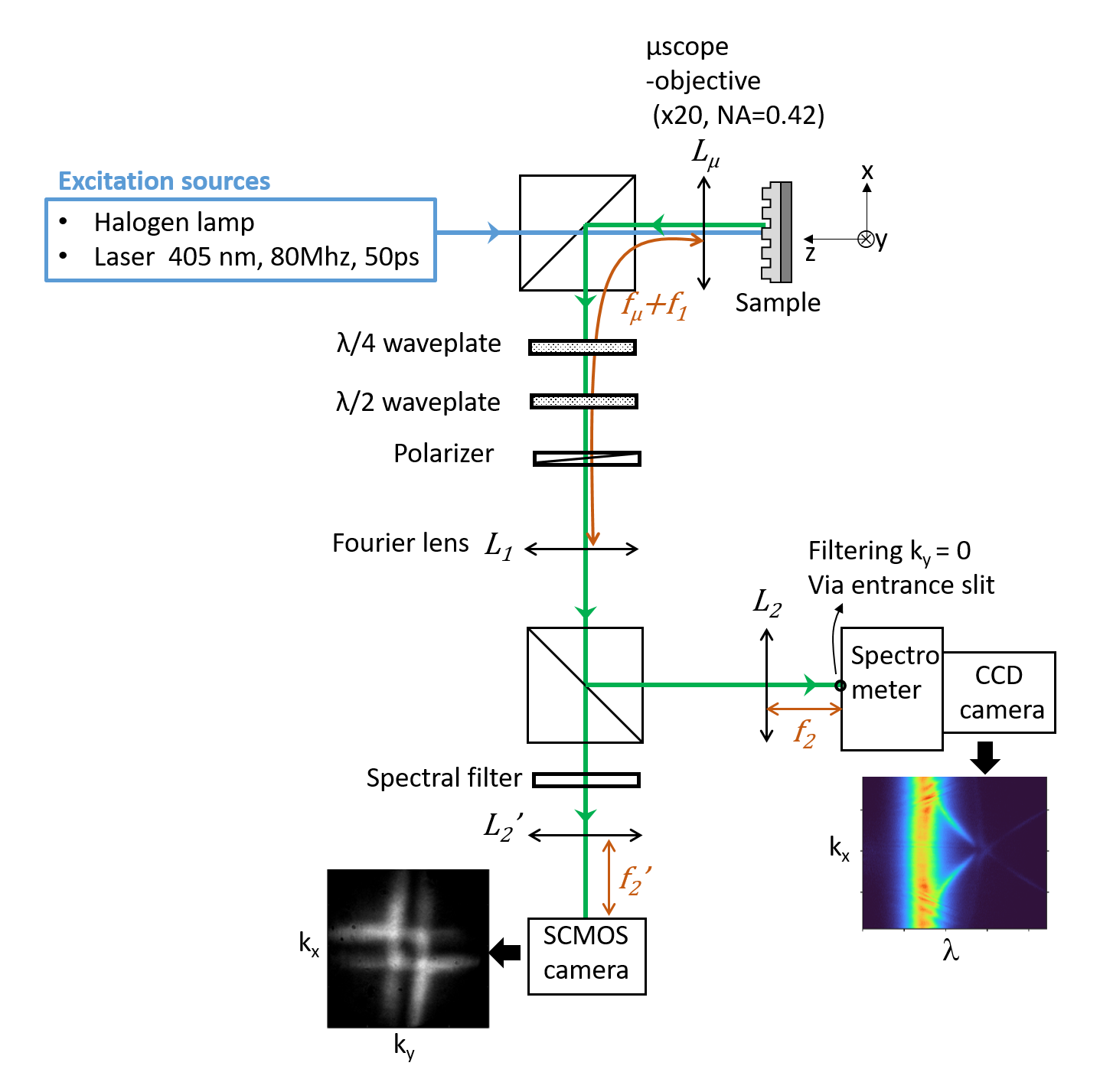}
		\end{center}
		\caption{\label{figS6}Optical characterization setup for Fourier spectroscopy.}
	\end{figure*}
	
\end{document}